\documentclass{emulateapj}
\usepackage{apjfonts}

\slugcomment{To appear in the Astronomical Journal}
\shorttitle{An All-Sky Catalog of Bright M Dwarfs.}

\begin{document}

\title{An All-Sky Catalog of Bright M Dwarfs.}

\author{S\'ebastien L\'epine\altaffilmark{1}, and Eric
  Gaidos\altaffilmark{2}}

\altaffiltext{1}{Department of Astrophysics, Division of Physical Sciences,
American Museum of Natural History, Central Park West at 79th Street,
New York, NY 10024, USA}

\altaffiltext{2}{Department of Geology \& Geophysics, University of
  Hawai'i, 1680 East-West Road, Honolulu, HI 96822, USA}

\begin{abstract}
We present an all-sky catalog of M dwarf stars with apparent infrared
magnitude J$<$10. The 8,889 stars are selected from
the ongoing SUPERBLINK survey of stars with proper motion $\mu>$40 mas
yr$^{-1}$, supplemented on the bright end with the TYCHO-2
catalog. Completeness tests which account for kinematic
(proper motion) bias suggest that our catalog represents $\approx75\%$
of the estimated $\sim11,900$ M dwarfs with J$<$10 expected to
populate the entire sky. Our catalog is, however, significantly more
complete for the Northern sky ($\approx$90\%) than it is for the South
($\approx$60\%). Stars are identified as cool, red M dwarfs
from a combination of optical and infrared color cuts, and are
distinguished from background M giants and highly-reddened stars using
either existing parallax measurements or, if such measurements are
lacking, on their location in an optical-to-infrared reduced
proper motion diagram. These bright M dwarfs are all prime targets for
exoplanet surveys using the Doppler radial velocity or transit
methods; the combination of low-mass and bright apparent magnitude
should make possible the detection of Earth-size planets on
short-period orbits using currently available techniques. Parallax
measurements, when available, and photometric distance estimates are
provided for all stars, and these place most systems within 60 parsecs
of the Sun. Spectral type estimated from V-J color shows that most of the
stars range from K7 to M4, with only a few late M dwarfs, all within 20
pc. Proximity to the Sun also makes these stars good targets for
high-resolution exoplanet imaging searches, especially if younger
objects can be identified on the basis of X-ray or UV excess. For that
purpose, we include X-ray flux from ROSAT and FUV/NUV ultraviolet
magnitudes from GALEX for all stars for which a counterpart can be
identified in those catalogs. Additional photometric data include
optical magnitudes from Digitized Sky Survey plates, and infrared
magnitudes from 2MASS.
\end{abstract}



\section{Introduction}

M dwarfs are main sequence stars whose spectra display bands of TiO
and other molecules such as CaH, VO, FeH, and CrH
\citep{Kirkpatrick.etal.1991}. The high molecular opacities are the
result of a cool atmosphere with effective temperatures in the range
2400K-3700K \citep{Allard.Hauschildt.1995}. Observed orbital motion of
astrometric binaries \citep{Delfosse.etal.1999} show that M dwarfs
have masses ranging from $\approx$0.5 M$_{\odot}$ all the way
down to the hydrogen-burning limit ($\approx$0.075 M$_{\odot}$). M
dwarfs have relatively small sizes, with radii roughly proportional to
their masses. Recent measurements of eclipsing systems
\citep{Fernandez.etal.2009} indicate a roughly linear
relationship between masses and radii, with a 0.4 M$_{\odot}$ (0.1
M$_{\odot}$) M dwarf having a radius of 0.4 R$_{\odot}$ (0.15
R$_{\odot}$). By number, M dwarfs form the bulk of the stars in our
Galaxy. The census of stars within 33 parsecs of the Sun
\citep{Lepine.2005a} shows M dwarfs outnumbering all other
hydrogen-burning objects by a factor of three to one. Because of their
low mass-to-light ratio, M dwarfs contribute only a small fraction of
the light emission in galaxies, but they may be the dominant baryonic
component in at least some galaxies \citep{vanDokkum.Conroy.2010}.

In recent years there has been an increasing interest in M dwarfs due
to the discovery that they too host exoplanets. As of January
2011, 25\% of the 41 Doppler-confirmed planets with $M\sin i
<$30M$_{\oplus}$ are orbiting M dwarfs\footnote{from
  http://exoplanet.eu}. The Geneva/HARPS planet search team reports
that $>$30\% of low-mass stars harbor planets with masses
$<$30$M_{\earth}$ on orbits shorter than 50~d
\citep{2009A+A...507..487M} and preliminary {\it Kepler} results
suggest that M dwarfs host just as many, if not more, super-Earths and
Neptunes as solar-mass stars \citep{2011ApJ...736...19B}. Exoplanet
candidates have been identified by {\it Kepler} in stars having colors
consistent with M dwarfs \citep{2011ApJ...728..117B}, despite the fact
that very few M dwarfs are being targeted by the mission
\citep{Batalha.etal.2010}.

Large exoplanet surveys have now started to monitor sizable numbers of M
dwarfs, such as the M2K program which is targeting some 1,600 M dwarfs
for radial velocity
monitoring \citep{Apps.etal.2010}, and the MEarth project
\citep{Irwin.etal.2009,Irwin.etal.2010} which is designed to detect
exoplanet transits in nearby late-type M dwarfs.

The two principal methods of exoplanet detection, Doppler measurement
of radial velocity (RV) and photometric detection of transits, are 
more sensitive to planets around stars of lower mass. For a given planet
mass and orbital period, RV amplitude scales as $M_*^{-2/3}$.
In contrast to late-type M stars, the vast majority of early M dwarfs
are not rapid rotators or chromospherically active (W$_{H\alpha}>1$\AA)
\citep{Bochanski.etal.2005,West.etal.2008,Reiners.Basri.2008}
and are thus amenable to these search techniques. In fact, inactive K
and early M stars have weaker P-mode oscillations and lower Doppler
noise (``jitter'') than their F or G counterparts
\citep{Wright.2005,Isaacson.Fischer.2011}. Photometric transit
depth scales approximately as $M_*^{-2}$ for planets of similar
size. Because of their significantly smaller radii, M dwarfs display
deeper transits which do not require high precision photometry for
detection, as they do for G stars \citep{Nutzman.Charbonneau.2008}.

Another method of planet detection is high-contrast imaging at
infrared wavelengths where a (giant) planet is self-luminous
\citep{Lowrance.etal.2005,Chauvin.etal.2006,Lafreniere.etal.2007,Biller.2007,Kasper.etal.2007,Beichman.etal.2010,Heinze.etal.2010,Liu.etal.2010}. M
dwarf systems present a much more favorable planet/star contrast
ratio compared to solar-mass stars, due to the significantly lower
luminosity of M dwarfs. Young systems are of particular
interest because the luminosity of giant planets is predicted to be
orders of magnitude higher at ages of $10^6-10^8$~yr
\citep{Fortney.Nettelman.2010} providing a more favorable contrast,
even as the host star itself is moderately more luminous at this younger
age. Young solar mass stars can be identified by their enhanced
rotation, dynamo activity, and chromospheric emission
\citep{Kiraga.Stepien.2007}, although the trend for late M stars for
which the mechanism of dynamo operation is thought to be different is less clear
\citep{West.etal.2009,West.Basri.2009,Browning.etal.2010,Reiners.Basri.2010}.
Another challenge for this approach is the apparent rarity of giant
planets on long-period orbits around solar-mass stars
\citep{Ida.Lin.2004,Nielsen.Close.2010} combined with the
paucity of giant planets around M dwarfs compared to G dwarfs
\citep{Johnson.etal.2010}. It is also possible that young giant
planets may be less luminous and more difficult to detect than
previously predicted \citep{Fortney.etal.2008}.

Low luminosity M dwarfs are also attractive targets because planets
within their diminutive habitable zones are easier to detect
\citep{Kasting.Whitmire.Reynolds.1993,Scalo.etal.2007,Tarter.etal.2007,Gaidos.etal.2007}. The
correlation between high
metallicity and the presence of giant planets found among solar-mass
stars \citep{Gonzalez.1997,Fischer.Valenti.2005} has been recently extended to M dwarf stars
\citep{Johnson.Apps.2009,Johnson.etal.2010}, and metal-rich M dwarfs,
if they can be identified
\citep{Bonfils.etal.2005,Schlaufman.Laughlin.2010,Rojas-Ayala.2010},
would be especially propitious targets for planet searches.

Despite the attractiveness of M dwarfs for exoplanet surveys, there
still is no systematic, all-sky census of these objects to a useful
limiting magnitude. Part of the problem is that the
low-mass M dwarfs are stars of relatively low luminosities
(M$_V\gtrsim12$). All but the closest M dwarfs are too faint (V$>$12)
to be part of the {\it Hipparcos} catalog \citep{Perryman.etal.1997},
which means that existing M dwarf samples are selected
from proper motion catalogs with fainter magnitude limits. These
include the Tycho-2 catalogue \citep{Hog.etal.2000}, which extends the
Hipparcos catalog a few magnitudes fainter, and also the older and
deeper Luyten Half Second (LHS) and New Luyten Two-Tenths (NLTT)
catalogs \citep{Luyten.1979a,Luyten.1979b} which were the primary
source material for the well-known {\em Catalog of Nearby
  Stars} \citep{Gliese.Jahreiss.1991}. Newer and more complete proper
motion catalogs, such as the LSPM-North \citep{Lepine.Shara.2005} have
the potential to expand the list of available M dwarf targets. 

In this paper, we present the first all-sky catalog of bright (J$<$10)
M dwarfs. The sample is extracted from the all-sky SUPERBLINK proper
motion catalog of stars with $\mu>$40 mas yr$^{-1}$, an extension of
LSPM-North currently in development. The selection method is
detailed in \S2. The catalog itself is presented in \S3. The
completeness of the catalog is evaluated in \S4. Conclusions follow in
\S5.

\section{Selection method}

\subsection{Source: proper motion stars from SUPERBLINK}

\begin{figure}[tb]
\epsscale{1.15}
\plotone{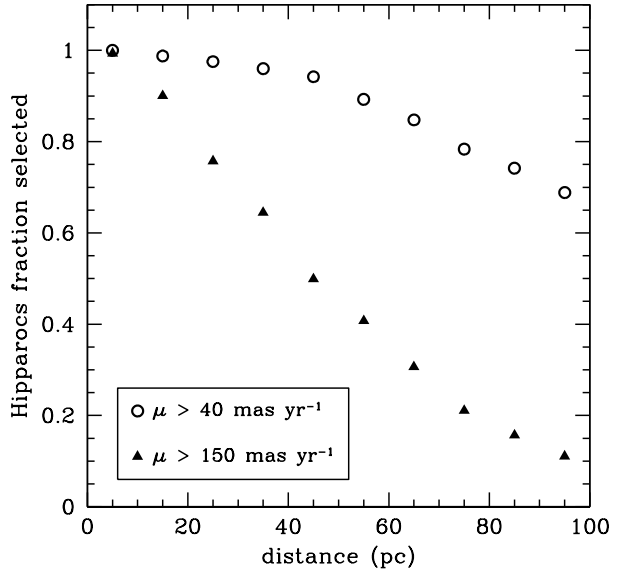}
\caption{Kinematic selection effects, illustrated by the fraction of
  nearby stars in the Hipparcos catalog which would be selected with a
  proper motion  cut $\mu>150$ mas yr$^{-1}$ (triangles) and $\mu>40$
  mas yr$^{-1}$  (circles) if placed at a specified distance. The
  fraction is calculated for stars in
  different   distance bins from 5$\pm$5pc to 95$\pm$5pc. While a
  $\mu>150$ mas   yr$^{-1}$ cut results in a heavy kinematic bias
  beyond 20pc, a   $\mu>40$ mas yr$^{-1}$ selects more than 80\% of
  stars all the way   to d$\simeq$75pc.}
\end{figure}

\begin{figure}[tb]
\epsscale{1.15}
\plotone{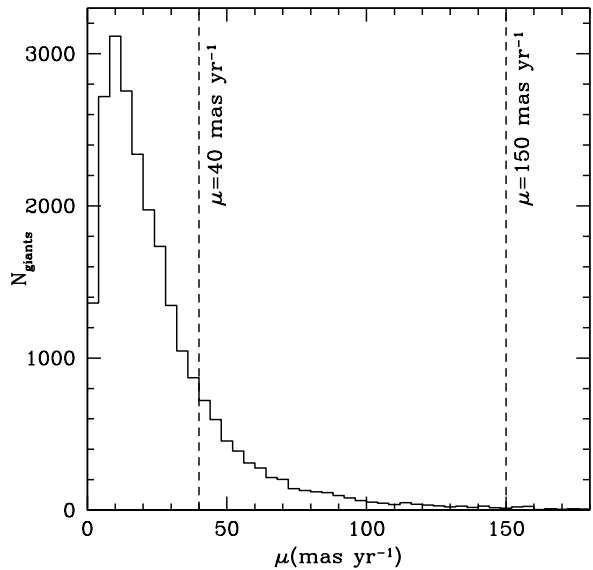}
\caption{Distribution of proper motions for red giants in the
  Hipparcos catalog. Due to their typically larger distances, most
  giants have proper motions below the SUPERBLINK limit ($\mu>40$ mas
  yr$^{-1}$). This minimizes the contamination of our M dwarf sample by
  background M giants.}
\end{figure}

The SUPERBLINK proper motion survey is an all-sky search for stars
with large proper motions based on a re-analysis of images from the
Digitized Sky Surveys (DSS) using a specialized image-differencing
algorithm and software. The basic search method is described in
\citet{Lepine.etal.2002}. Quality control procedures, including
cross-correlation with other catalogs and the compilation of
astrometric and photometric results, is discussed at length in
\citet{Lepine.Shara.2005}. Stringent quality control procedures,
including visual confirmation of most objects, guarantees a very low
level of false detections. The survey has a fixed, low proper motion
threshold of $\mu>40$ mas yr$^{-1}$. New stars with very large proper
motions ($\mu>450$ mas yr$^{-1}$) discovered in the survey have been
published in \citet{Lepine.etal.2002}, \citet{Lepine.etal.2003},
\citet{Lepine.2005b}, and \citet{Lepine.2008}. A complete list of
northern stars with proper motions $\mu>150$ mas yr$^{-1}$ was
published in \citet{Lepine.Shara.2005}; a similar list covering the
southern sky has been completed and is available from the
authors. Those complete lists incorporate data from the Hipparcos and
Tycho-2 catalogs at the bright end, and the catalogs are estimated to
be $>90\%$ complete to a visual magnitude $V=19.0$, and extend to
$V\simeq20.0$. The completeness is significantly higher at high
galactic latitudes, where field crowding is less of an issue. Dense
fields at low galactic latitudes have a lower completeness due
to faint stars often being lost to non-linearity in the crowded areas
of photographic plates; saturation also prevents the detection of
proper motion sources in the extended glare (up to several arc minutes)
of extremely bright stars (e.g. Vega, Sirius, Canopus).

As of July 2011, the full SUPERBLINK catalog (stars with proper
motions $\mu>40$ mas yr$^{-1}$) comprised 2,270,481 objects, with the
most extensive sky coverage north of Decl.=-20$^{\circ}$. The catalog
was complete for the northern sky (Decl.$>$0$^{\circ}$), but the survey
was still in progress for large parts of the southern sky. The
southern survey was however complete for stars with proper motions
$\mu>150$ mas yr$^{-1}$. This version of the SUPERBLINK catalog is the
one used here.

\subsection{Fundamental kinematic bias}

\begin{figure}[tb]
\epsscale{1.15}
\plotone{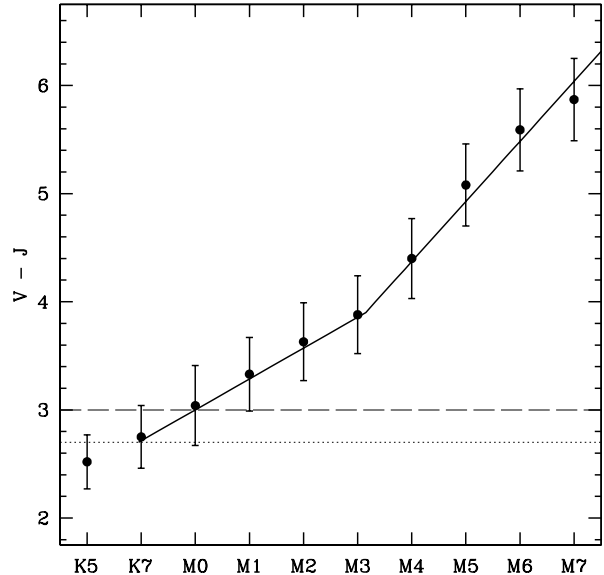}
\caption{Optical-to-infrared V$-$J colors of late-K and M dwarfs, as a
  function of spectral subtype. The values are averages from
  subsamples of M dwarfs from the SUPERBLINK proper motion survey
  which also have spectral classification from the Sloan Digital Sky
  Survey; error bars show the 1-$\sigma$ dispersion in each
  bin. A color cut $V-J>3.0$ (dashed line) yields a strict selection of
  M dwarfs with few late-K objects, while a $V-J>2.7$ cut (dotted
  line) is more inclusive for early M star at the expense of some
  late-K contamination. The latter color cut is adopted for our
  catalog. The solid line shows our adopted color-spectral type
  relationship for estimating subtypes based on photometry alone
  (\S3.5, and Equation 15).}
\end{figure}

\begin{figure*}[th]
\epsscale{1.0}
\plotone{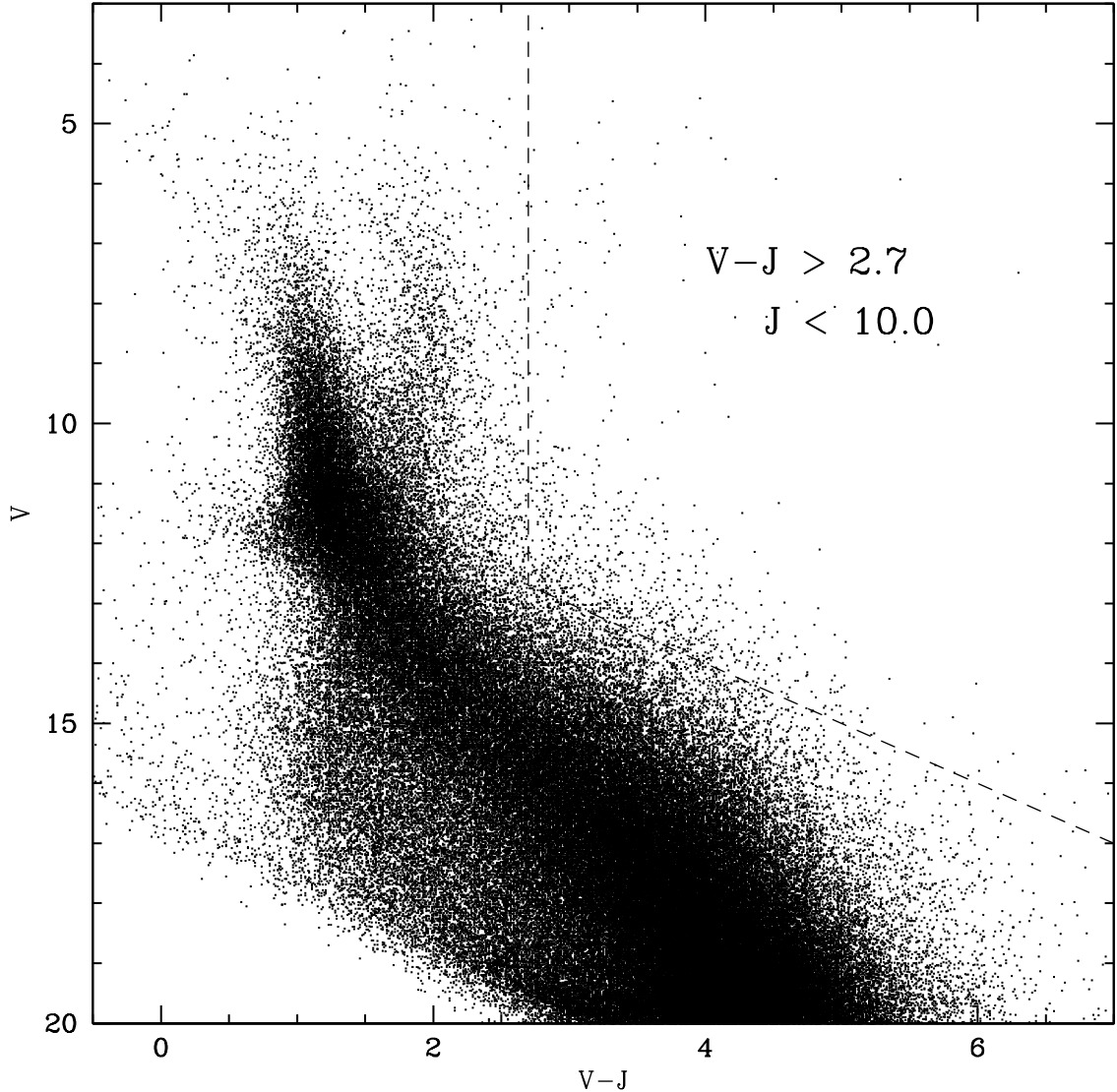}
\caption{Apparent magnitude vs. optical-to-infrared
  ($V-J$) color for a representative subset of 100,000 stars from the
  SUPERBLINK proper motion catalog (L\'epine et al. 2011, in
  preparation). From the full 2,270,481 SUPERBLINK sample, our
  selection box for bright 
  M dwarfs is shown ($V-J>2.7$; $J<10$); this selects 10,659 stars
  representing the brightest objects at any color beyond
  $V-J=2.7$. Only 1 in 230 stars from the full SUPERBLINK catalog are
  thus selected.} 
\end{figure*}

The use of a proper motion catalog is a convenient way to
identify nearby stars in general, and M dwarfs in particular, because
proper motion-selected samples discriminate against distant background
sources including M giants that
would otherwise significantly contaminate a sample of cool stars selected
by color alone. However, proper motion selection carries its own
inherent bias: stars with small velocities in the plane of the sky
tend not to be detected in proper motion surveys, depending on the
distance to the source. Nearby, slow-moving stars can be included by
lowering the proper motion limit of the sample, but this has the
effect of increasing the contamination from background sources. There
is a fine balance to achieve in adjusting the proper motion threshold,
which would ideally have the sample incorporate the largest possible
fraction of stars up to a specified distance, minimizing the kinematic
bias, while at the same time limiting the contamination from background
objects. 

The SUPERBLINK catalog, with a proper motion limit $\mu>40$
mas yr$^{-1}$, achieves such a balance for stars located within 100
parsecs of the Sun. This can be demonstrated with an analysis of
proper motion in the {\it Hipparcos} catalog. Figure 1
shows the fraction of Hipparcos stars at a given distance from the Sun
which have proper motions $\mu>150$ mas yr$^{-1}$ and $\mu>40$ mas
yr$^{-1}$. The former value is the proper motion limit of the
LSPM-North and south catalogs (which is also the approximate proper
motion limit of the NLTT catalog and, by extension, the Catalog of
Nearby Stars), while the latter is the proper motion limit of the
extended SUPERBLINK catalog. The diagram shows that operating with a
proper motion limit of $\mu>150$ mas yr$^{-1}$ will only detect half
the stars at 40pc and very few stars (only those with very large
components of motion) at 100pc. However a sample with a proper motion
limit $\mu>40$ mas yr$^{-1}$ will include $\approx$95\% of the stars
at 40pc and $\approx$70\% of the stars at 100pc.

Background giant stars, on the other hand, tend to have small proper
motions, which exclude them from the SUPERBLINK catalog. Figure 2
shows the distribution of proper motions for red giant stars in the
Hipparcos catalog, selected with the color and absolute magnitude cuts
$B-V>$1.0, $M_V<$3.0. The bulk of the giants have proper motions below
the SUPERBLINK threshold. This particular subset of giants, because it
is from the Hipparcos catalog, also represents the nearest red giants
to the Sun, since they all have magnitudes $V<12$. More distant giants
would have even smaller proper motions, and would also be excluded
from our sample.

\subsection{Color and magnitude cuts for bright M stars}

Figure 3 shows the distribution of
optical-to-infrared $V-J$ colors for a subset of M dwarfs with spectral
classification from the Sloan Digital Sky Survey database
\citep{West.etal.2011}. Infrared $J$ magnitudes are from 2MASS, while
$V$ magnitudes are the SUPERBLINK visual magnitudes, estimated from
Palomar Sky Survey photographic plate measurements as described in
\citet{Lepine.Shara.2005}. The subset includes all stars in SDSS which
have counterparts in the SUPERBLINK proper motion catalog. For this subset
of nearby M dwarfs, $V-J$ color ranges from $V-J\simeq2.7$ at subtype
M0, to $V-J\simeq8.0$ at subtype M9. We adopt an M dwarf color-cut
of $V-J>2.7$, which is expected to include most stars of subtype M0
and later, at the expense of some contamination from late-K stars.

In addition, we select stars that are bright enough to be monitored
by current and planned Doppler radial velocity instruments. Stars as
faint as $V=12$ are routinely monitored using the High Resolution
Spectrometer (HIRES) on the 10~m Keck I Telescope,
e.g. \citet{Johnson.etal.2007,Johnson.etal.2010b}, and fainter but
highly meritorious stars with transiting planets, e.g. GJ 1214
\citep{Charbonneau.etal.2009}, have been observed. M dwarf stars are
brighter in the near-infrared, and future high-resolution
spectrometers should be able to detect planets around stars as faint
as $J=10$. Anticipating such developments, we select stars with
$J<10$, which also includes all M dwarfs with $V<12.7$.

To summarize, we select bright M dwarfs among high proper motion stars
($\mu>40$ mas yr$^{-1}$) in the SUPERBLINK catalog using these
color/magnitude cuts:
\begin{eqnarray}
V-J > 2.7 \\
J < 10.0 
\end{eqnarray}

Figure 4 shows the full magnitude-color distribution of SUPERBLINK
stars, and the region targeted by our selection. Out of 2,270,481
stars in the SUPERBLINK catalog, our color and magnitude cuts select
10,659 bright red stars. This we use as our initial sample to identify
bright M dwarfs.

The sky distribution of the color-selected stars is shown in Figure
5. Separate plots show the stars listed in the TYCHO-2 and HIPPARCOS
catalogs (top), stars from the SUPERBLINK catalog which are not
listed in TYCHO-2 or HIPPARCOS {\em and} have proper motions $\mu>$
150 mas yr$^{-1}$ (center), and finally stars from the SUPERBLINK
catalog not listed in TYCHO-2 or HIPPARCOS {\em and} with proper
motions 40 mas yr$^{-1}<\mu<$ 150 mas yr$^{-1}$ (bottom). The
subsample from the TYCHO-2 and HIPPARCOS catalogs includes 3,319 of
the brighter stars, spread over the entire sky. Their distribution
shows a slight overdensity in the direction of the Galactic bulge,
which is likely due to the larger number of red giants found in that
part of the sky. The very high proper motion ($\mu>$ 150 mas
yr$^{-1}$) subsample from SUPERBLINK includes 4,398 stars, and extends
uniformly over the entire sky. The moderately high proper motion (40
mas yr$^{-1}<\mu<$ 150 mas yr$^{-1}$) subsample from SUPERBLINK
includes 2,941 stars, and clearly suffers from incompleteness in the
southern hemisphere; the subsample also displays an overdensity around
($\alpha$,$\delta$)$\approx$(80,+20) which is associated with the
nearby Hyades cluster.

\begin{figure}[t]
\epsscale{2.15}
\plotone{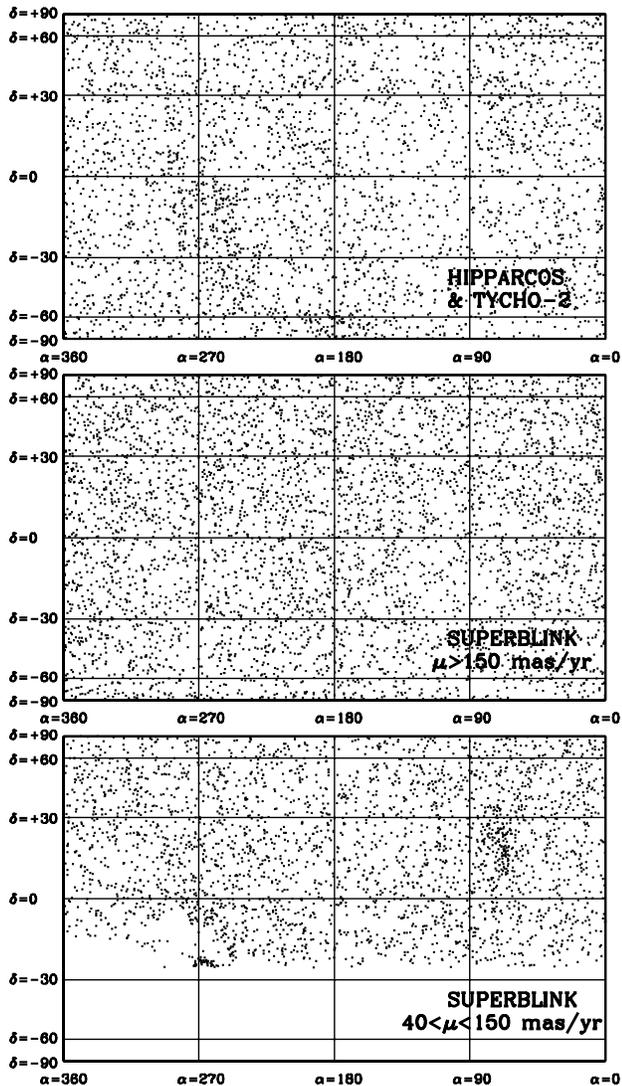}
\caption{Sky distribution of the various proper motion surveys used in
  building our catalog, in a Gall-Peters equal-area projection. Top:
  bright M dwarfs from the Hipparcos and Tycho-2 catalogs (3319
  stars). Center: M dwarfs identified from the all-sky SUPERBLINK
  survey with proper motion limit $\mu>150$ mas yr${^-1}$ (4398
  stars). Bottom: M dwarfs identified from the extension of the
  SUPERBLINK survey to proper motions 40 mas yr${^-1}<\mu<$150 mas
  yr${^-1}$ (2941 stars). The first two samples cover the entire sky
  while the latter covers only part of the southern declinations.}
\end{figure}

\subsection{Elimination of background giants}

Giant stars constitute the main source of contaminant in the
sample. Indeed, an excess of red sources in the apparent magnitude
range $5<V<10$ (see Figure 4) points to a modest population of very
red M giants in the SUPERBLINK catalog (probably including AGB
stars). While these giants are only a tiny fraction of the full
SUPERBLINK catalog, they are systematically selected in our initial
color/magnitude cut. To eliminate these giants from our sample we
apply three selection criteria. The first is based on evaluation of
absolute magnitudes, and is used on stars with existing parallax
data. The second method identifies giants from their location in a
reduced proper motion diagram, and is used for stars for which
there are no existing parallax measurements. A third method selects
for M dwarfs based on $J-H$ and $H-K_s$ colors, further eliminating
giants and interlopers with bad optical and/or infrared photometry.

\subsubsection{Stars with parallax: absolute magnitude cut}

\begin{figure}[tb]
\epsscale{1.15}
\plotone{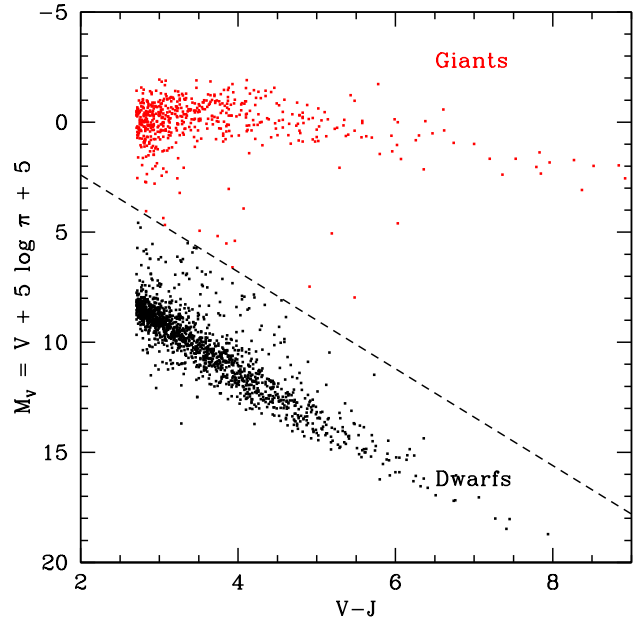}
\caption{Color-absolute magnitude diagram for candidates with measured
  trigonometric parallaxes, a total of 2,080 stars. Red giants are
  distinguished by their significantly brighter absolute
  magnitudes. We define a demarcation line to separate dwarfs from
  giants. This identifies 561 red giants (plotted in red) leaving
  1,519 probable red dwarfs (plotted in black).}
\end{figure}

We searched the literature to find astrometric parallaxes for stars
in our subsample of M dwarf candidates. Parallaxes for 1,622 stars
were recovered from the updated version of the Hipparcos catalog
\citep{2007A+A...474..653V}. Ground-based astrometric parallaxes were
additionally recovered for 10 stars in \citet{2011AJ....141..117J}, 46
stars in \citet{2010AJ....140..897R}, 15 stars in
\citet{2010AstL...36..576K}, 6 stars in \citet{2009AJ....137.4109L}, 3
stars in \citet{2009AJ....137..402G}, 1 star in
\citet{2008AJ....136..452G}, 2 stars in \citet{2007A+A...464..787S},
18 stars in \citet{2006AJ....132.2360H}, 4 stars in
\citet{2005AJ....130..337C}, 14 stars in \citet{2005AJ....129.1954J},
15 stars in \citet{1995GCTP..C......0V}, 3 stars in
\citet{1993AJ....105.1571H}, and 3 stars in
\citet{1992AJ....103..638M}. Whenever parallaxes were found for one
star in more than one bibliographical source, we adopted the most
recent measurement, following the order above. The numbers quoted
above refer to incremental additions to the list of parallaxes. In
addition, we found parallaxes for 3 stars in the
SKY2000 Master Catalog, Version 4 \citep{SKY2000}, and for 315 more
from the NStars
database\footnote{http://nstars.nau.edu/nau\_nstars/index.htm}. In all
cases we only retained parallax measurements with errors $<$20\%. The
final tally comprises a total of 2,080 stars with astrometric
parallaxes.

From the parallax $\pi$ we calculate absolute visual magnitudes using:
\begin{equation}
M_V = V + 5 \log{\pi} + 5.
\end{equation}
The resulting Hertzsprung-Russel (H-R) diagram is displayed in Figure
6. The stars populate two very distinct loci. The more luminous group
includes all M giants and the less luminous group consists of all the
M dwarfs. For stars with $V-J\simeq2.7$, the giants are typically 7-8
magnitudes more luminous than the dwarfs. The difference in luminosity
only increases at redder colors, with the reddest giants almost 20
magnitudes more luminous than dwarfs of similar $V-J$ color. 

From the absolute magnitude $M_V$ and optical-to-infrared color $V-J$,
we define the following criterion for a star to be considered an M
dwarf:
\begin{equation}
M_V > 2.2 ( V-J ) - 2.0 .  
\end{equation}
This cut identifies 561 stars as most likely to be red giants. The
remaining 1,519 objects are identified as M dwarfs. We emphasize that
this selection works only for the $\approx20\%$ stars in our initial
sample for which astrometric parallax measurements are
available. Note that the color-magnitude relationship for the M dwarfs
in our sample closely follows:
\begin{equation}
M_V \approx 2.2 ( V-J ) + 2.5 ,  
\end{equation}
which implies that we are eliminating stars that are more than 4.5
magnitudes overluminous compared to the average M dwarf of the same
color. This cut could possibly eliminate some very young M dwarfs,
which tend to be overluminous in an optical-to-infrared
color-magnitude diagram \citep{Hawley.etal.1999}. Indeed a number of
stars with parallaxes appear to be a few magnitudes over-luminous
compared to average field stars. However a close examination of Figure
6 indicates that our magnitude cut would, in the worst case, eliminate
only a handful of objects that could have been overluminous dwarfs.

\subsubsection{Stars without parallax: reduced proper motion cut}

\begin{figure}[tb]
\epsscale{2.2}
\plotone{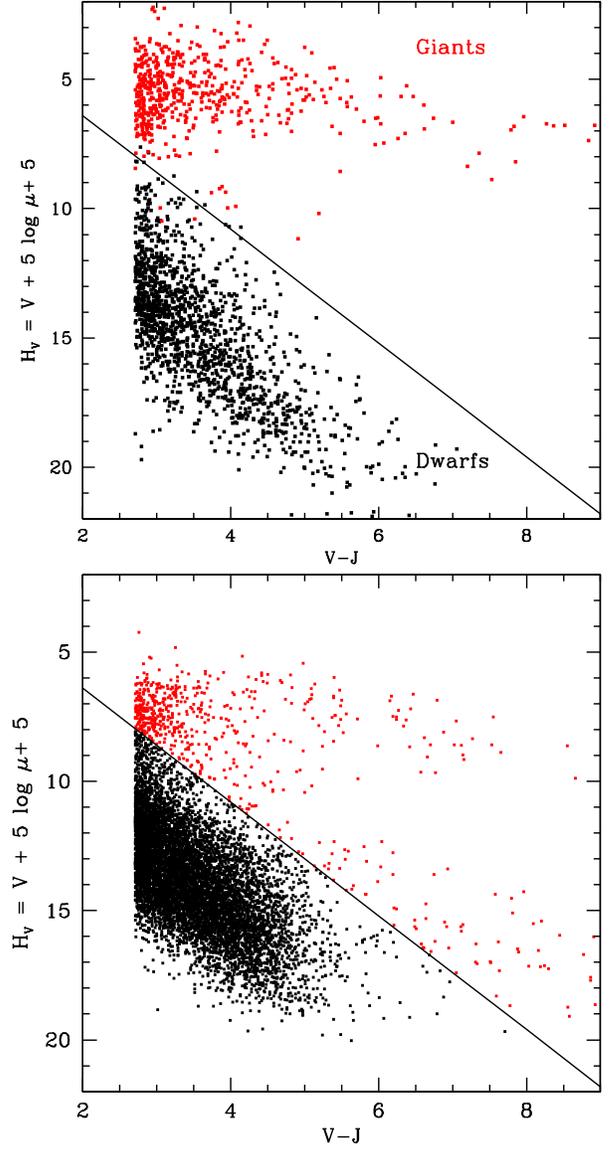}
\caption{Reduced proper motion diagrams for stars in our original
  sample. Top: diagram for stars with astrometric parallax
  measurements (see Fig.6), with the M giants plotted in red (561
  objects) and the dwarfs plotted in black (1,519 objects). The line
  shows our adopted demarcation between giant stars and dwarfs. The
  dashed line shows our proposed demarcation line. Bottom: diagram for
  candidates with no parallax. Based on the same demarcation line, we
  exclude 626 objects as probable M giants. The selection leaves
  7,954 candidate M dwarfs, all plotted in black.}
\end{figure}

For stars without parallax measurements, it is possible to identify
giants based on their location in a reduced proper motion diagram. The
reduced proper motion is defined as:
\begin{equation}
H_V = V + 5 \log{\mu} + 5, 
\end{equation}
a quantity which is analogous to the absolute magnitude $M_V$ (Eq.3)
except that the proper motion $\mu$ is substituted for the parallax
$\pi$. The two quantities are mathematically connected, with:
\begin{equation}
H_V \equiv M_V + 5 \log{v_T} - 3.38 ,
\end{equation}
where $v_T$ is the projected motion in the plane of the sky, i.e. the
transverse velocity, here expressed in km s$^{-1}$. This means that
the reduced proper motion diagram is basically an H-R diagram where the
ordinates are modulated by the individual components of transverse
velocities.
 
Just as in the H-R diagram, M giants and M dwarfs tend to occupy
distinct loci, with the giants all having relatively low values of
$H_V$ and clustering in the upper part of the diagram. We demonstrate
this in the upper panel of Figure 7, which shows the reduced
proper motion diagram of the stars in our sample for which there are
parallax measurements. Stars identified as giants based on their
parallax are plotted in red, dwarfs are plotted in black. We define a
demarcation line which separates stars of the two groups quite
well. From this, we postulate that M dwarfs can be separated from
giants following this criterion:
\begin{equation}
H_V > 2.2 ( V-J ) + 2.0 .
\end{equation}
Because the reduced proper motion depends on both absolute
magnitude {\em and} transverse velocity, our selection may eliminate
some main sequence M dwarfs if their transverse motion is small. Based
on equations 5 and 7, we see that stars on the mean main sequence will
be eliminated if their transverse velocity $V_T\lesssim3.77$
km s$^{-1}$. Given that the velocity dispersion for stars in the Solar
Neighborhood is $\approx$40 km s$^{-1}$ \citep{Nordstrom.etal.2004},
our adopted criterion will exclude relatively few M dwarfs.

Conversely, giants with large transverse velocities will also have
large values of $H_V$ which may place them beyond the limit defined in
Equation 8. Among the parallax-confirmed giants, we find 4 stars which
fall within our defined reduced proper motion cut for M dwarfs. One of
the stars has a inferred transverse velocity $V_T=206$ km s$^{-1}$
based on its Hipparcos parallax and proper motion. However the other
three stars have suspiciously faint absolute magnitudes
(4.5$<M_V<$5.0) for M giants, and may instead be M dwarfs with
erroneous (underestimated) parallaxes measurements.


The reduced proper motion diagram of stars with no parallax
measurements is shown in the bottom panel of Figure 7. Of the 8,580
stars with proper motions but no parallax measurements, we identify
626 as probable M giants, and those are eliminated from the
sample. The remaining 7,954 stars are retained as M dwarf
candidates. This leaves 9,473 stars, with or without parallax, on the
list of M bright M dwarf candidates.

A significant number of sources are found to have very red colors
($V-J>$5) and reduced proper motion in the range typical for dwarf
stars ($H_V>$12) and yet fall above the line we defined as the
giant-dwarf boundary. A close examination of their infrared colors
shows that most of these stars fall outside the normal range for M
dwarfs (see below) and are likely either stars with erroneous
(overestimated) V magnitudes, or distant F-G-K stars with large
components of reddening. Most of these stars are eliminated after the
infrared cuts described below.

\subsubsection{Additional color-color cuts}
 
Giant stars are notable for having infrared colors different from the
M dwarfs, most apparent in a [($J-H$),($H-K$)] color-color diagram. This
has been known since \citet{Bessell.Brett.1988}. Dwarf stars have
comparatively bluer $J-H$ colors than giants. This is due to opacity
differences, mostly from molecular bands of H$_2$O, which yield
significantly higher $H$-band and $K$-band opacities in M dwarfs
\citep{Bessell.Castelli.Plez.1998}.

\begin{figure}[tb]
\epsscale{2.2}
\plotone{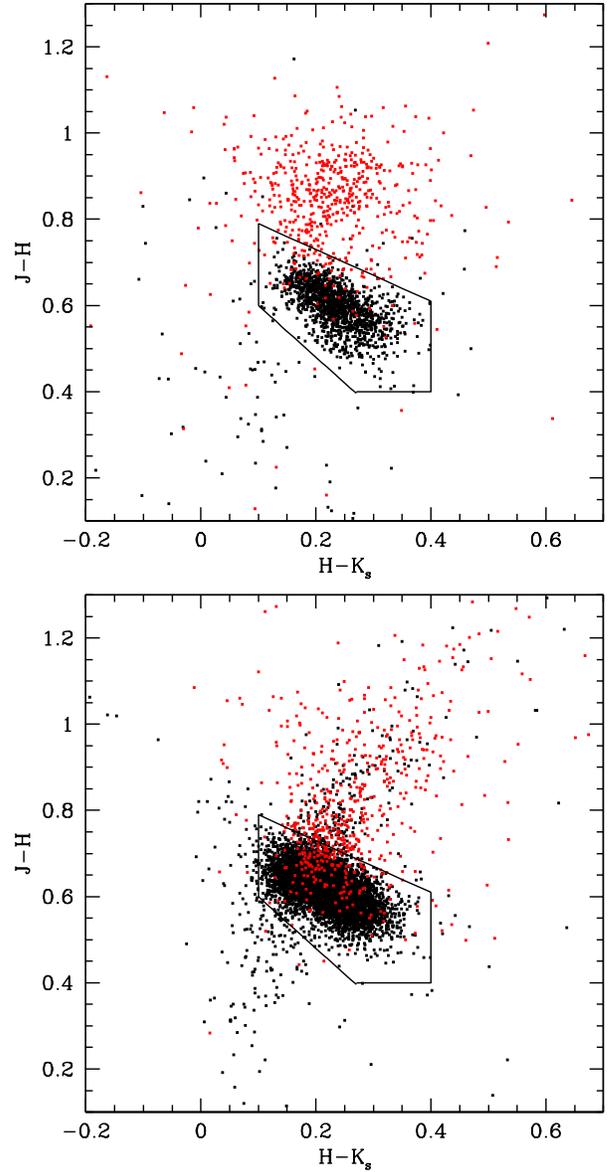}
\caption{Infrared color-color diagram for the candidate objects. Top:
  diagram for stars with astrometric parallaxes (see Fig.6), with the
  giants plotted in red (561 objects) and the dwarfs plotted in black
  (1,519 objects). Dwarfs and giants have distinct loci, with the giants
  having generally redder $J-H$ values. We define a box where
  most of the M dwarf stars are located. Objects outside the box are
  more likely to be M giants, or stars with erroneous photometric
  data. Bottom: diagram for the 8,580 stars with no parallax. Some 942
  objects do not fall within our defined ``red dwarf box'' and are
  eliminated from the sample.}
\end{figure}

Figure 8 shows the [($J-H$),($H-K$)] color-color distribution for the
stars in our initial sample. Stars with parallax measurements are
shown in the upper panel, stars without parallax in the lower
panel. In both panels, all objects we have identified as probable
giants, based on either absolute magnitude (\S2.4.1) or reduced proper
motion(\S2.4.2), are plotted in red. The remaining M dwarf candidates
are plotted in black. The giants are clearly segregated from the
dwarfs along J-H, although there is a small overlap between the two
groups. In fact M dwarfs appear to be confined to a relatively compact
region around  $J-H=0.6$ and $H-K=0.25$. A small number of alleged M
dwarfs fall outside the cluster; some have red $J-H$ colors more
consistent with the M giants, and indeed they most probably are. Other
stars have bluer colors which are more typical of F-G-K stars; we
suspect these probably are more massive main sequence stars which made
it into our sample probably due to an overestimated photographic
magnitude $V$, which would also overestimate the $V-J$ color term,
bringing them into the M star subsample. In order to eliminate these
leftover misidentified M
giants and misclassified F-G-K stars, we further restrict our M dwarf
selection to the region in [($J-H$),($H-K$)] diagram bounded by:
\begin{eqnarray}
J-H < 0.85 - 0.6 ( H-K_s )\\
J-H > 0.72 - 1.2 ( H-K_s )\\
J-H > 0.40\\
H-K_s > 0.10\\ 
H-K_s < 0.40
\end{eqnarray}
Among stars with trigonometric parallaxes, we find that 97 of the M
dwarf candidates (black dots in Figure 8 upper panel) fall outside the
limits of the region defined above. We conclude that these stars
likely have significant errors in their estimated optical or infrared
magnitudes, and we exclude them from our census of M dwarfs. 

Among stars with no parallax measurements identified as probable M
dwarfs from the reduced proper motion cut (black dots in Figure 8), we
find 487 which fall outside the defined M dwarf boundaries; these
stars are eliminated as well. Some of the stars have color consistent
with F-G-K main sequence stars (0$<H-K_s<$0.2, 0.1$<J-H<$0.6) and most
likely made the initial $V-J>$2.7 color cut due to an underestimated V
magnitude. Other excluded sources have $J-H>$0.7, and are either
giants or distant main sequence stars with large reddening. After
these cuts, we are left with a list of 8,889 probable
bright M dwarfs, which we retain for building our catalog.

\subsubsection{Giant contamination vs M dwarf completeness}

Among the giants confirmed through parallax (stars plotted in red in
the upper panel of Figure 8) a small number fall within the selection
box for M dwarfs. Specifically, 66 of the 561 giants fall within the
region defined by Equations 9-13. This means that JHK color cuts alone
are susceptible to contamination by background giants. This makes the
reduced proper motion selection critical in minimizing giant
contamination. Note that only 4 of the 561 giants have reduced proper
motions within the limit defined for M dwarfs, which increases our
confidence that giant contamination in our catalog is negligible.

The reduced proper motion cut, on the other hand, may be eliminating
bona fide M dwarfs, as mentioned in \S2.4.2 above. The lower panel in
Figure 8 show evidence for this, as 160 of the 626 stars excluded
after the reduced proper motion cut (plotted in red) do fall within
the M dwarf box. Overall this suggests that our adopted selection
criteria places more of an emphasis on avoiding contamination by
giants, at the expense of some incompleteness in the M dwarf
targets. From the numbers above, we predict that $\sim$100 M dwarfs
might be missing from our catalog because they fail the reduced proper
motion cut.

In addition, we note that stars identified as giants based on parallax
data (Figure 8 top panel, red dots) and stars excluded from the
reduced proper motion cuts (Figure 8 bottom panel, red dots) show
significantly different distributions in $J-H$/$H-K$ color-color
space. One average the parallax giants tend to have bluer $H-K_s$
colors. We suggest that this is due to the fact that most red giants
with $\mu>$40 mas yr$^{-1}$ are bright enough to have parallaxes from
the Hipparcos catalog, and that the reduced proper motion cut for
stars with no parallax essentially serves to exclude stars with large
components of reddening, which are expected to have redder $H-K_s$
colors than the giants.

\begin{deluxetable*}{lllrrrrrrrc}
\scriptsize
\tablecolumns{11}
\tablewidth{0pc}
\tablecaption{Bright M dwarfs, designations and
  astrometry\tablenotemark{1}}
\tablehead{
\colhead{SUPERBLINK \#} &
\colhead{Hipparcos \#} &
\colhead{TYCHO-2 \#} &
\colhead{CNS3 name} &
\colhead{$\alpha$} &
\colhead{$\delta$} &
\colhead{$\mu$}&
\colhead{$\mu_{\alpha}$} &
\colhead{$\mu_{\delta}$} &
\colhead{$\pi$} &
\colhead{src\tablenotemark{1}} \\
\colhead{} &
\colhead{} &
\colhead{} &
\colhead{} &
\colhead{(ICRS)} &
\colhead{(ICRS)} &
\colhead{mas yr$^{-1}$} &
\colhead{mas yr$^{-1}$} &
\colhead{mas yr$^{-1}$} &
\colhead{mas} &
\colhead{} 
}
\startdata
PM I00001$+$6943  & \nodata&         \nodata &  \nodata&   0.028531&  69.717117&  0.136&   0.136&  -0.002& \nodata& \nodata\\ 
PM I00003$-$0802S & \nodata&         \nodata &  \nodata&   0.087698&  -8.037146&  0.100&   0.029&  -0.096& \nodata& \nodata\\ 
PM I00005$-$0533  & \nodata&         \nodata &  \nodata&   0.144895&  -5.551960&  0.193&   0.181&   0.067& \nodata& \nodata\\
PM I00006$+$1829  & \nodata& TYC 1181-1683-1 &  \nodata&   0.163528&  18.488850&  0.387&   0.335&   0.195& \nodata& \nodata\\
PM I00007$+$1624  & \nodata&         \nodata &  \nodata&   0.195877&  16.402805&  0.125&   0.013&  -0.125& \nodata& \nodata\\
PM I00007$-$3510  & \nodata&         \nodata &  \nodata&   0.195306& -35.168343&  0.316&   0.312&  -0.048& \nodata& \nodata\\
PM I00007$-$6243  & \nodata&         \nodata &  \nodata&   0.189068& -62.728802&  0.199&   0.166&  -0.110& \nodata& \nodata\\
PM I00012$+$1358S & \nodata& TYC  600-1507-2 &  \nodata&   0.303578&  13.972055&  0.147&   0.025&   0.144&  24.9$\pm$1.0& VA95\\
PM I00014$+$4724  & \nodata&         \nodata & GJ 1293 &   0.371546&  47.414665&  0.172&   0.171&  -0.008& \nodata& \nodata\\
PM I00014$-$1656  & HIP 112&  TYC 5838-784-1 &  \nodata&   0.357633& -16.948410&  0.393&   0.299&  -0.255&  31.4$\pm$4.2& VL07\\
PM I00015$-$0814  & \nodata&         \nodata &  \nodata&   0.399365&  -8.244875&  0.124&   0.097&  -0.077& \nodata& \nodata\\
PM I00016$-$3258  & \nodata&         \nodata &  \nodata&   0.408577& -32.978249&  0.238&   0.238&   0.005& \nodata& \nodata\\
PM I00016$-$7613  & \nodata&         \nodata &  \nodata&   0.408879& -76.230522&  0.188&   0.188&   0.008& \nodata& \nodata\\
PM I00017$-$3528  & \nodata&         \nodata &  \nodata&   0.433436& -35.476082&  0.502&   0.502&  -0.023& \nodata& \nodata\\
PM I00024$+$0440  & \nodata&         \nodata &  \nodata&   0.613839&   4.668418&  0.122&   0.058&   0.105& \nodata& \nodata\\
PM I00024$-$4601  & HIP 191&  TYC 8022-637-1 &  \nodata&   0.612922& -46.028893&  0.194&   0.192&  -0.022&  23.0$\pm$4.4& VL07\\
PM I00025$-$6324  & \nodata&         \nodata &  \nodata&   0.639251& -63.401974&  0.822&  -0.488&  -0.661& \nodata& \nodata\\
PM I00026$+$3821  & \nodata&         \nodata &  \nodata&   0.667151&  38.362587&  0.075&  -0.073&  -0.020& \nodata& \nodata\\
PM I00026$-$1123  & \nodata&         \nodata &  \nodata&   0.654399& -11.393147&  0.584&  -0.413&  -0.413& \nodata& \nodata\\
PM I00026$-$3919  & \nodata&         \nodata &  \nodata&   0.654961& -39.320053&  0.200&   0.199&   0.023& \nodata& \nodata\\
\enddata
\tablenotetext{1}{The full version of this table is available in the
  electronic version of the Astronomical Journal. The first twenty
  lines of the table are printed here to show the general layout.}
\tablenotetext{2}{Source of the astrometric parallax: 
  Co05 = \citet{2005AJ....130..337C},
  Ga08 = \citet{2008AJ....136..452G},
  Ga09 = \citet{2009AJ....137..402G},
  Ha93 = \citet{1993AJ....105.1571H},
  He06 = \citet{2006AJ....132.2360H},
  Ja05 = \citet{2005AJ....129.1954J},
  Ja11 = \citet{2011AJ....141..117J},
  Kh10 = \citet{2010AstL...36..576K},
  Le09 = \citet{2009AJ....137.4109L},
  Mo92 = \citet{1992AJ....103..638M},
  My02 = \citet{SKY2000},
  Ri10 = \citet{2010AJ....140..897R},
  Sm07 = \citet{2007A+A...464..787S},
  VA95 = \citet{1995GCTP..C......0V},
  VL07 = \citet{2007A+A...474..653V},
  NSTA = NStars database (http://nstars.nau.edu/nau\_nstars/index.htm)}
\end{deluxetable*}

\section{The catalog}

\subsection{Source Identifications}

Each of the 8,889 stars in our catalog is identified with the standard
proper motion star ID used in the SUPERBLINK proper motion
survey. These SUPERBLINK names are tabulate in the first column of
both Table 1 and Table 2. The first three letters (``PMI'') are the
SUPERBLINK catalog
identifier. These are followed by 5 digits which relate to the ICRS
right ascension of the star in sexagesimal; the first four digits are
the hours and minutes of right ascension, the fifth digit is the
seconds of right ascension divided by 6 and rounded down to the
nearest integer. These are followed by the declination sign and then
four digits which replicate the ICRS declination (hours, then minutes)
of the source. A disambiguation letter (NSEW) is used when the scheme
would leave two stars with the same name, with the letter indicating
the relative orientation of the stars in the pair. This scheme follows
the recommendation of \citet{Eggen.1979} and is the convention now
adopted in reporting results from the SUPERBLINK survey, see
e.g. \citet{Lepine.2005b,Lepine.2008}.

In Table 1, We additionally provide the names of the stars in both the
Hipparcos (column 2) and TYCHO-2 (column 3) catalogs, for all stars which have
counterparts in those catalogs. These are provided for convenience
because Hipparcos names are now commonly used to refer to very bright
stars, and because many exoplanet surveys have adopted the Hipparcos
names in designating the host stars and their exoplanets. The TYCHO-2
names perhaps do not enjoy the same popularity, but are useful in
identifying objects for which the proper motion is known to higher
accuracy. Our list includes a total of 977 Hipparcos and 1,859
TYCHO-2 stars. Some 779 entries have counterparts in both the
Hipparcos and TYCHO-2 catalogs, but 192 stars have Hipparcos
counterparts only, while 1,074 stars are only listed in TYCHO-2.

In addition, we identify by their catalog names all the stars listed
in the Third Catalog of Nearby Stars \citep{Gliese.Jahreiss.1991} or
CNS3 (column 4). These are also convenient because CNS3 names have
been widely adopted by the exoplanet community to designate nearby
stars. The CNS3 counterparts are found after cross-correlation with
the catalog of revised positions by \citet{Stauffer.etal.2011} which
lists 4,106 objects. A total of 1,529 stars from our catalog of bright
M dwarfs are thus matched to their CNS3 counterpart.

Stars with CNS3 counterparts include most of the very nearest
systems. However, the CNS3 does suffer from a significant kinematic
bias because it is largely based on the proper motion catalogs of
Luyten, particularly the LHS catalog which has a relatively high
proper motion limit $\mu\gtrsim$450 mas yr$^{-1}$. As a result, many
stars on our list which we identify as very nearby objects do
not have CNS3 names; these are typically stars with relatively small
($\mu\lesssim300$ mas yr$^{-1}$) proper motions.

We note that $67.8\%$ of the stars in our catalog
(6,003 entries) have no counterpart in either the Hipparcos, TYCHO-2,
or CNS3 catalog. Use of the PM designations for these stars would be
convenient, especially given that the PM name indicates the location
of the star on the sky.

\subsection{Astrometric data}

\subsubsection{Coordinates}

Coordinates in the table are tabulated in columns 5 and 6. Right
ascension ($\alpha$) and declination ($\delta$) are listed in the ICRS
system, and quoted for the 2000.0 epoch. This means that current
positions must be extrapolated using the tabulated proper motions. For
stars listed in the Hipparcos catalog, we have extrapolated the
positions to the 2000.0 epoch, from the values listed in
\citet{2007A+A...474..653V} which were listed for epoch 1991.25. For all
other stars, the 2000.0 coordinates are based on the position of their
2MASS counterparts as listed in \citet{Cutri.etal.2003} and
extrapolated to the 2000.0 epoch from the epoch of the 2MASS
observations. The positions are typically accurate to 0.8$\arcsec$,
which are the quoted 2MASS catalog errors on the absolute astrometry;
Proper motion errors have little effect on the accuracy of the
extrapolated positions because of the proximity of the 2MASS survey
epoch to the millennium year ($<3$ years).

\subsubsection{Proper motions}

The total proper motion $\mu$ of the stars are tabulated in column 7 and
listed in seconds of arc per year. The vector components of the proper
motion in the directions of R.A. and
Decl. ($\mu_{\alpha},\mu_{\delta}$) are tabulated in columns 8 and
9. Both components are given in seconds of arc per year, following the
SUPERBLINK convention. Note that $\mu_{\alpha}$ is listed in seconds
of arc per year, not seconds of time per year, by which we mean that
$\mu_{\alpha}\equiv\dot{\alpha}\sin{(\delta)}$.

These proper motions are from three separate sources. Stars with
Hipparcos counterparts are listed with their proper motion from the
Hipparcos catalog \citep{2007A+A...474..653V}. Stars not in the Hipparcos
catalog but with counterparts in the TYCHO-2 catalog are listed with
their proper motion from TYCHO-2 \citep{Hog.etal.2000}. The Hipparcos
and TYCHO-2 proper motions are typically accurate to 0.5-2.0 mas
yr$^{-1}$. For stars not listed in either the HIPPARCOS or TYCHO-2,
the proper motions listed are those measured in the SUPERBLINK proper
motion survey. Those proper motions are based on a re-analysis of the
Digitized Sky Surveys images, as described in detail in
\citet{Lepine.Shara.2005}; these proper motions have a typical
precision of $\pm$8 mas yr$^{-1}$. One can determine the source of
the proper motion by checking whether the star has a Hipparcos or
TYCHO-2 name.

\subsubsection{Parallaxes}

Astrometric parallaxes are recovered from the literature for 1,422 of
the M dwarfs in our catalog. See section 2.4.1 for the list of papers
and catalogs that were searched to recover the parallaxes. All values
are listed in column 10 whenever available, otherwise the field is
left blank. A flag is added in column 11 to indicate the source of the
parallax, with the key provided in the table footnote. In all cases,
measurement errors as reported in the relevant source, are noted in
column 10.

\begin{deluxetable*}{lrrrrrrrrrrrrrrc}
\scriptsize
\tablecolumns{16}
\tablewidth{0pt}
\tablecaption{Bright M dwarfs, photometry, photometric distances, and
  estimated subtypes.\tablenotemark{1}}
\tablehead{
\colhead{SUPERBLINK \#} &
\colhead{X-ray\tablenotemark{2}} &
\colhead{FUV\tablenotemark{3}} &
\colhead{NUV} &
\colhead{B$_{\rm T}$\tablenotemark{4}} &
\colhead{V$_{\rm T}$} &
\colhead{B$_{\rm J}$\tablenotemark{5}} &
\colhead{R$_{\rm F}$} &
\colhead{I$_{\rm N}$} &
\colhead{J\tablenotemark{6}}&
\colhead{H} &
\colhead{K$_{\rm s}$} &
\colhead{V\tablenotemark{7}} &
\colhead{V$-$J} &
\colhead{$\pi_{\rm phot}$\tablenotemark{8}} &
\colhead{subtype\tablenotemark{9}} \\
\colhead{} &
\colhead{cts s$^{-1}$} &
\colhead{mag} &
\colhead{mag} &
\colhead{mag} &
\colhead{mag} &
\colhead{mag} &
\colhead{mag} &
\colhead{mag} &
\colhead{mag} &
\colhead{mag} &
\colhead{mag} &
\colhead{mag} &
\colhead{mag} &
\colhead{mas} &
\colhead{} \\
}
\startdata
PM I00001$+$6943  &  0.0461&\nodata&\nodata&\nodata&\nodata& 14.3& 12.6& 10.8&  9.70&  9.12&  8.84&  13.52&   3.82& 33.1$\pm$09.9 & m3   \\
PM I00003$-$0802S & \nodata&\nodata&\nodata&\nodata&\nodata& 13.9& 11.8&\nodata& 9.12&  8.47&  8.27&  12.93&   3.81& 42.9$\pm$12.9 & m3   \\
PM I00005$-$0533  & \nodata&\nodata&\nodata&\nodata&\nodata& 13.2& 11.2& 10.2&  9.00&  8.39&  8.17&  12.28&   3.28& 31.3$\pm$09.4 & m1   \\
PM I00006$+$1829  & \nodata&\nodata&\nodata& 12.91& 11.28& 11.9& 10.3&  9.5&  8.44&  7.79&  7.64&  11.28&     2.84& 30.7$\pm$09.2 & k7   \\
PM I00007$+$1624  & \nodata&\nodata& 21.02&\nodata&\nodata& 14.0& 11.9&\nodata&  9.32&  8.68&  8.46&  13.03&  3.71& 16.1$\pm$04.8 & m0   \\
PM I00007$-$3510  & \nodata&\nodata&\nodata&\nodata&\nodata& 13.3& 11.1& 10.2&  9.12&  8.48&  8.28&  12.29&   3.17& 27.4$\pm$08.2 & m1   \\
PM I00007$-$6243  & \nodata&\nodata&\nodata&\nodata&\nodata& 13.7& 11.7& 11.0&  9.89&  9.23&  9.07&  12.78&   2.89& 36.5$\pm$10.9 & m2   \\
PM I00012$+$1358S & \nodata&\nodata& 19.85& 13.67& 11.12&\nodata&\nodata&\nodata&  8.36&  7.71&  7.53& 11.12& 2.76& 30.6$\pm$09.2 & k7   \\
PM I00014$+$4724  & \nodata&\nodata&\nodata&\nodata&\nodata& 13.8& 11.4& 10.6&  9.67&  9.02&  8.83&  12.70&   3.03& 35.4$\pm$10.6 & k7   \\
PM I00014$-$1656  & \nodata&\nodata&\nodata& 12.14& 10.76& 11.5&  9.8&  8.9&  8.02&  7.41&  7.22&  10.76&     2.74& 19.3$\pm$05.8 & m0   \\
PM I00015$-$0814  & \nodata&\nodata&\nodata&\nodata&\nodata& 14.4& 11.7&\nodata&  9.79&  9.12&  8.91&  13.16& 3.37& 23.1$\pm$06.9 & m1   \\
PM I00016$-$3258  & \nodata&\nodata&\nodata&\nodata&\nodata& 13.4& 11.6& 10.7&  9.77&  9.14&  8.94&  12.57&   2.80& 16.3$\pm$04.9 & k7   \\
PM I00016$-$7613  & \nodata&\nodata&\nodata&\nodata&\nodata& 14.1& 12.6& 10.6&  9.48&  8.89&  8.60&  13.41&   3.93& 39.5$\pm$11.9 & m3   \\
PM I00017$-$3528  & \nodata&\nodata&\nodata& 14.99& 13.36& 14.5& 12.1& 11.0&  9.82&  9.19&  8.93&  13.36&     3.54& 25.7$\pm$07.7 & m2   \\
PM I00024$+$0440  & \nodata&\nodata& 21.96&\nodata&\nodata& 13.6& 11.2&  9.9&  9.19&  8.61&  8.45&  12.50&    3.31& 28.2$\pm$08.5 & m1   \\
PM I00024$-$4601  & \nodata&\nodata&\nodata& 13.92& 12.43&\nodata&\nodata&\nodata&  9.18&  8.52& 8.34& 12.43& 3.25& 29.3$\pm$08.8 & m1   \\
PM I00025$-$6324  & \nodata&\nodata&\nodata&\nodata&\nodata& 14.2& 12.1& 10.7&  9.32&  8.71&  8.53&  13.23&   3.91& 41.9$\pm$12.6 & m3   \\
PM I00026$+$3821  & \nodata& 21.97& 20.15&\nodata&\nodata& 14.8& 13.1& 10.6&  9.71&  9.20&  8.91&  14.02&     4.31& 20.3$\pm$06.1 & m1   \\
PM I00026$-$1123  & \nodata&\nodata&\nodata& 14.74& 13.09&\nodata&\nodata&\nodata & 9.86& 9.24& 9.03& 13.09&  3.23& 17.1$\pm$05.1 & m0   \\
PM I00026$-$3919  & \nodata&\nodata&\nodata&\nodata&\nodata& 14.0& 11.4& 10.9&  9.84&  9.19&  9.00&  12.80&   2.96& 45.3$\pm$13.6 & m4   \\
\enddata
\tablenotetext{1}{The full version of this table is available in the
  electronic version of the Astronomical Journal. The first twenty
  lines of the table are printed here to show the general layout.}
\tablenotetext{2}{X-ray flux from the ROSAT all-sky point source
  catalog \citep{Cutri.etal.2003}.}
\tablenotetext{3}{Far-UV and near-UV magnitudes from the GALEX 5th
  data release.}
\tablenotetext{4}{B$_T$ and V$_T$ magnitudes from the TYCHO-2 catalog.}
\tablenotetext{5}{Photographic B$_{\rm J}$ (blue IIIa J), R$_{\rm F}$
  (red IIIa F), and infrared I$_{\rm N}$ (infrared IV N) magnitudes from
  USNO-B1.0 catalog \citep{Monet.etal.2003}.}
\tablenotetext{6}{Infrared J, H, and K$_{\rm s}$ magnitudes from the
  2MASS catalog \citep{Cutri.etal.2003}.}
\tablenotetext{7}{Effective V magnitude as defined in \citet{Lepine.Shara.2005}.}
\tablenotetext{8}{Photometric parallax based on the ($M_V$,$V-J$)
  relationship defined in this paper.}
\tablenotetext{9}{Estimated spectral subtypes based on the (ST,$V-J$)
  relationship defined in this paper.}
\end{deluxetable*}

\subsection{Photometric data}

Table 2 lists photometric data for the stars, along with photometric
parallaxes and estimated spectral types, both based on the photometric
measurements. SUPERBLINK stars names are repeated in column 1 of Table
2, to facilitate the cross-identification with Table 1, although the
number of lines are the same.

\subsubsection{X-ray}

We have cross-correlated our list of bright M dwarfs with both the
{\it ROSAT} All-Sky Bright Source Catalog \citep{Voges.etal.1999}
and the {\it ROSAT} All-Sky Survey Faint Source Catalog
\citep{Voges.etal.2000}. We used a search radius of 15\arcsec, which
is on the order of the astrometric precision of the ROSAT catalog. Our
search identified 1,065 stars with X-ray counterparts. Flux values in
counts s$^{-1}$ are listed in column 2.

X-ray fluxes can be useful in identifying young stars in the
catalog. M dwarfs can be X-ray emitters if they have significant
chromospheric activity. High activity levels are typically linked to
rapid rotation, which in M dwarfs is indicative of either spin-orbit
coupling in a close binary \citep{2006AJ....131.1674S} or of young age
\citep{2005AJ....129.2428S}. It is likely that the youngest M dwarfs
in our catalog are to be found among the subsample of objects with
significant X-ray emission. Follow-up observations of X-ray bright M
dwarfs has been successful in identifying scores of low-mass members
to young moving groups
\citep{Gaidos.1998,Zuckerman.etal.2001,Montes.etal.2001,Torres.etal.2006}.

\subsubsection{Ultra-violet}

The catalog was cross-correlated against the fifth data
release (DR5) of the GALEX mission. Using a $5\arcsec$ search radius,
we found GALEX counterparts for 3,905 of the 8,889 stars on our
list. The $NUV$ and $FUV$ magnitudes are listed in columns 3 and 4. Some
762 objects have counterparts in both the $FUV$ and $NUV$, while 3,115
have counterparts only in the $NUV$, leaving 28 stars with a counterpart
in the $FUV$ only. These ultra-violet (UV) magnitudes can also be used
to identify young stars. A high UV luminosity is associated with
activity in M dwarfs, and can be used as an age diagnostic
\citep{Shkolnik.etal.2011}.

\subsubsection{TYCHO-2 blue and visual magnitudes}

We include optical magnitudes $B_T$ and $V_T$ as they are listed in
the HIPPARCOS and TYCHO-2 catalogs. These are included in columns 5
and 6 respectively. These are available only for stars with
counterparts in those catalogs. The $B_T$ and $V_T$ magnitudes are
useful because they are generally more accurate (0.1 mag or better)
than the photographic magnitudes which we list for all the stars.

\subsubsection{Optical photographic}

Optical magnitudes are obtained from the USNO-B1.0 catalog of
\citet{Monet.etal.2003}, which is based on scans of historical
photographic surveys. The blue magnitudes are extracted mostly from
scans of $IIIaJ$ plates from the Palomar Sky Surveys (POSS-I, POSS-II)
and the Southern ESO Schmidt (SERC) Survey. Red magnitudes are
extracted from scans of $IIIaF$ plates from POSS-I and POSS-II and
also from the Anglo-Australian Observatory red survey
(AAO-red). Photographic infrared magnitudes are extracted from $IVn$
plates from POSS-II and SERC. Cross-correlation with the USNO-B1.0
catalog is performed as part of the SUPERBLINK quality control
procedure, and all ambiguous cases are verified by visual examination
using overlays of the USNO-B1.0 sources on the DSS scans. All the
recovered photographic magnitudes are listed in columns 7 (blue), 8
(red) and 9 (infrared).

Since the optical $b$, $r$, $i$, magnitudes are based on scans of
photographic plates, they are generally not as reliable as magnitudes
measured on electronic detectors, such as those from 2MASS and
GALEX. Photographic magnitudes are typically accurate to only $\pm0.5$
mag.

%

\subsubsection{Infrared}

As part of the quality control process, all the stars in the
SUPERBLINK proper motion survey have their counterparts identified in
the 2MASS catalog of \citet{Cutri.etal.2003}. As described in \S3.2.1
above, the 2MASS counterparts provide the absolute astrometry for
SUPERBLINK objects. Because of the color restriction of Equation 1,
every object in our catalog is required to have a counterpart in
2MASS. The corresponding infrared $J$, $H$, and $K_{\rm s}$ magnitudes
are listed in columns 10, 11, and 12, respectively. The magnitudes are
generally accurate to $\pm$0.08 mag, though one should refer to the
2MASS documentation for a detailed discussion of magnitude errors and
uncertainties\footnote{http://www.ipac.caltech.edu/2mass/releases/allsky/doc/explsup.html}. Very
bright stars ($J<5$), in particular, are saturated in 2MASS and have larger
photometric uncertainties.

\subsubsection{Effective visual magnitude}

For convenience, we calculate an effective visual magnitude $V$,
which for bright stars is the visual magnitude $V_T$ quoted in the
HIPPARCOS or TYCHO-2 catalogs. For faint stars, it is a combination
of the photographic $b$, $r$, and $i$ magnitudes, following the
algorithm described
in \citet{Lepine.Shara.2005} where this effective visual magnitude
is labeled as $V_e$. In this paper, we use the shorter form $V$ for
convenience, but it should be understood that all the mentions of $V$
actually refer to this effective visual magnitude $V_e$. All values
are listed in column 13.

\begin{figure}[t]
\epsscale{1.15}
\plotone{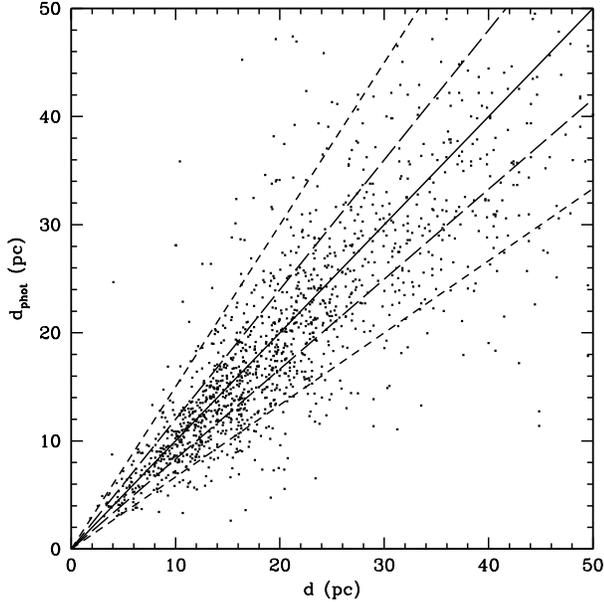}
\caption{Comparison between estimated photometric distances $d_{phot}$
  and the actual distances $d$ as measured from trigonometric
  parallaxes, for the stars in our catalog for which reliable parallax
  measurements do exist. The long-dashed lines show the extent of a
  $\pm$20$\%$ error on the photometric distance, the short-dashed line
  show the extent of a $\pm$50$\%$ error. Most stars fall
  within the latter limits.}
\end{figure}

\subsection{Photometric distances}

Photometric distances are estimated for all the stars based on the
color magnitude relationship for M dwarfs calibrated in
\citet{Lepine.2005a}. For stars with $V-J>2.7$ the relationship is:
\begin{equation}
M_V = \left[ 
\begin{array}{ccc}
2.09 ( V-J ) + 2.78 & \forall & 2.7<V-J<3.0 \\
2.52 ( V-J ) + 1.49 & \forall & 3.0<V-J<4.0 \\
2.35 ( V-J ) + 2.17 & \forall & 4.0<V-J<5.0 \\
1.89 ( V-J ) + 4.47 & \forall & V-J>5.0
\end{array}
\right].
\end{equation}
Photometric parallaxes $\pi_{phot}$ are then calculated based
on the estimated $M_V$ and the apparent visual magnitude $V$. Values
are listed in column 14.

As discussed in \citet{Lepine.2005a}, the relationship has a scatter
of 0.7 mag about the mean, which typically results in photometric
distances with errors of $\pm50\%$. In Figure 9, we
compare the photometric distances to triangulated distances, for stars
with available parallaxes. Of the 1,422 M dwarfs with parallaxes, some
1,206 stars have photometric distance estimates within $\pm50\%$ of
the astrometric distance, but only 772 have photometric distances
that are accurate to $\pm20\%$. There is also a significant number of
outliers which could be unrecognized giants, young field stars, or
unresolved doubles, all of which would have distances underestimated
by photometry. Metal-poor stars (M subdwarfs) on the other hand would
have their distances overestimated by photometry. Spectroscopic
follow-up would be required to identify all such objects.

Thus, while photometric distances are conveniently provided here for
all the stars, one should bear in mind that these are mainly for
guidance purposes, as these distances carry large uncertainties in
many cases. Astrometric parallaxes (listed in Table 1) should be
preferred for stars for which they are available, and should be
measured for the remaining objects in our catalog.

\begin{figure}[t]
\epsscale{1.15}
\plotone{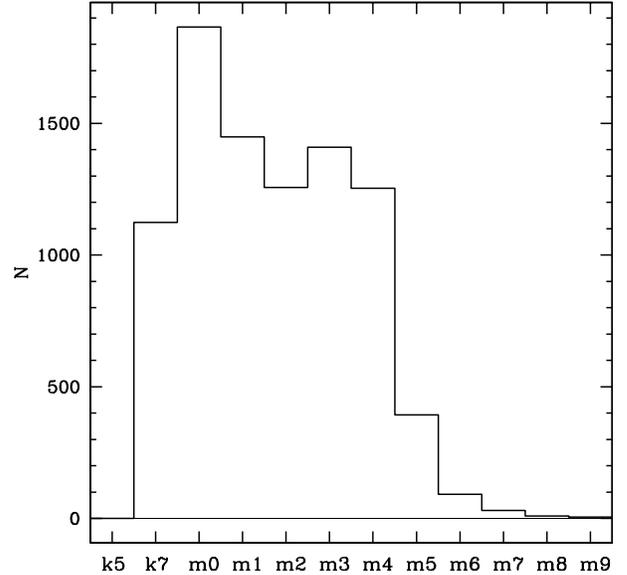}
\caption{Distribution of spectral subtypes as estimated from
  optical-to-infrared $V-J$ colors. These photometric subtypes are
  provided only for guidance, as they may be different from the formal
  spectral types. The photometric subtypes nevertheless  provide a
  reasonable assessment of the catalog contents. As expected for a
  sample of relatively bright M dwarfs, the catalog is mostly
  represented by early-type objects, which tend to be more luminous,
  and breaks down beyond spectral subtype M4.}
\end{figure}

\subsection{Estimated spectral type}

For all the stars, we provide an estimated spectral subtype which is
based on the relationship between $V-J$ color and the spectral subtype
(see Fig. 4). We use this simple two-component linear
relationship to estimate the subtype $ST$:
\begin{equation}
ST = \left[ 
\begin{array}{ccc}
3.50 ( V-J ) - 10.50 & \forall & V-J<3.9 \\
1.80 ( V-J ) - 3.87 & \forall & V-J\geq3.9
\end{array}
\right],
\end{equation}
after which the values are rounded up to the nearest integer. We use
lowercase letters (k,m) instead of the usual uppercase (K,M) to
indicate that the spectral type is only an estimate; a value of
$ST=-1$ yields an estimated subtype of k7. A histogram of the
estimated subtype distribution is shown in Figure 10, where it becomes
clear that our magnitude-limited catalog of bright M dwarfs is biased
toward stars of earlier subtypes. The initial mass function is
expected to peak at around 0.3 M$_{\odot}$ or spectral subtype M3-M4
\citep{Kroupa.etal.2002}, so our incompleteness starts at around M0
and becomes severe by M5.

These subtypes are tabulated in column 15 and may be used as a
guide in selecting targets of interest. Some of the stars on the list,
in particular stars listed in the CNS3, already have published
literature about them, and the interested user would be advised to
verify the existence of formal spectral classification(s). We are
currently conducting a spectroscopic follow-up survey to obtain formal
spectral types for the brightest stars in our catalog, and intend to
provide formal spectral classifications for these in the near future.

\section{Extent and completeness of the catalog}

\begin{figure*}[t]
\epsscale{1.10}
\plotone{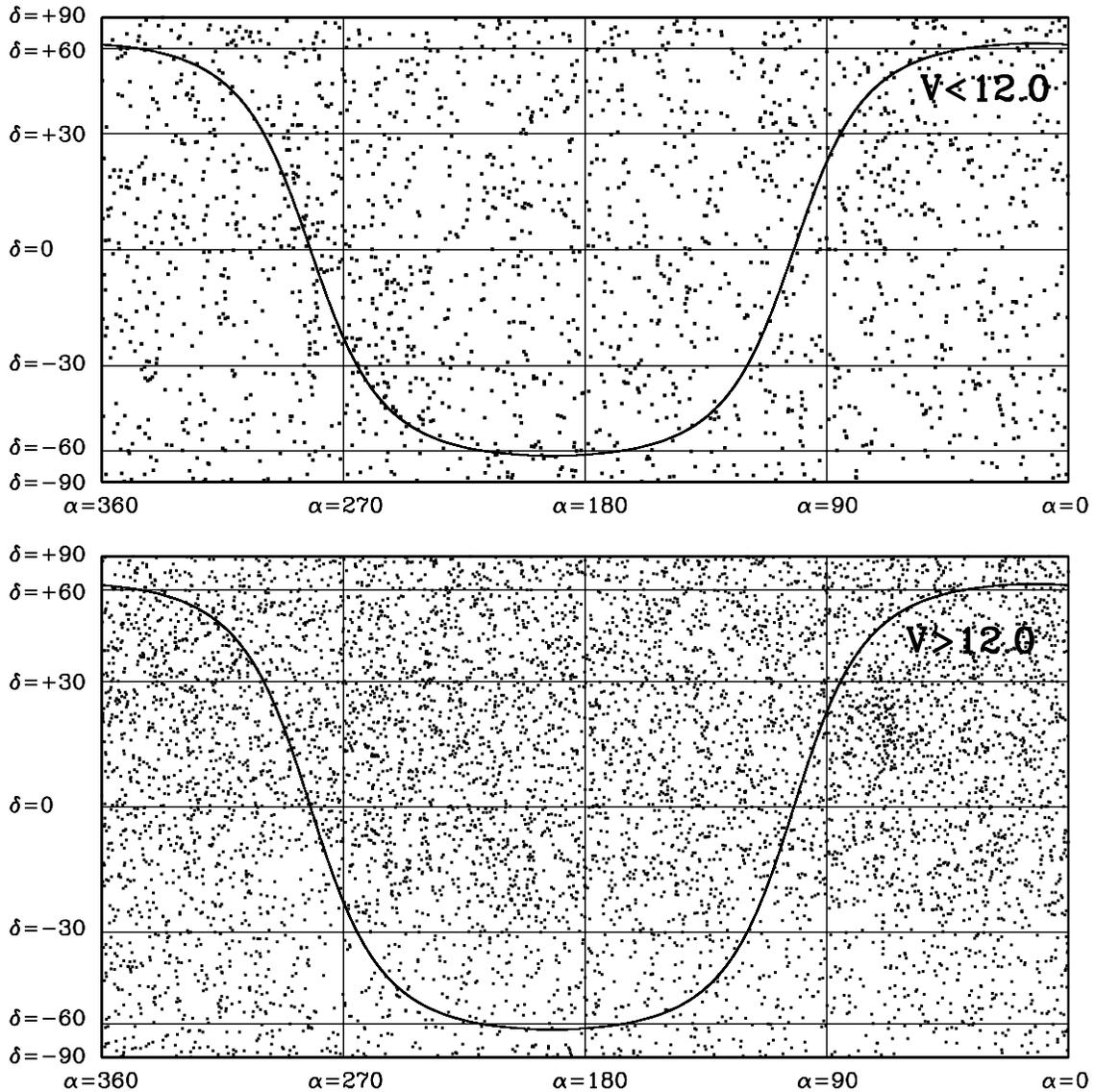}
\caption{Distribution on the sky of our catalogued bright M dwarfs, in
  a Gall-Peters equal-area projection. The top panels plots the 1,993
  stars in our catalog with relatively bright visual magnitudes
  ($V<12.0$), while the bottom panel shows the 6,896 fainter objects
  ($V>12.0$). The distribution of brighter M dwarfs is relatively
  uniform, but the fainter sources show a dearth of objects in the
  southern sky. This is a combination of the faint magnitude limit
  ($V\lesssim13$) of the TYCHO-2 catalog and the higher proper motion
  limit of the SUPERBLINK catalog ($\mu>150$ mas yr$^{-1}$) in most
  areas of the southern sky, while the northern sky extends all the
  way down to $\mu>40$ mas yr$^{-1}$ (see Figure 5).}
\end{figure*}

\subsection{Sky distribution}

Distribution on the sky of all the stars in our catalog is displayed
in Figure 11 in a Galls-Peter equal-area projection. To emphasize
differences between the brighter and fainter stars in our catalog, we
display the distribution in two panels, with one showing only stars
with visual magnitudes $V<12.0$ (upper panel) and the other showing
stars with visual magnitudes $V>12.0$. The separation roughly
corresponds to the magnitude limit of the TYCHO-2 catalog. Of the
1,993 stars in the brighter sample, 1,614 are listed in either the
TYCHO-2 or HIPPARCOS catalog. The fainter sample, on the other hand,
contains only 437 HIPPARCOS and TYCHO-2 stars out of a total of 6,896
objects.

The distribution of the brightest ($V<12$) M dwarfs appears relatively
uniform over the sky, except for an overdensity at low Galactic
latitude in the general direction of the Galactic center
($\alpha\sim270, \delta\sim-30$). This overdensity is likely due to
contamination by background giants in dense fields. Since our catalog
is dependent on the HIPPARCOS and TYCHO-2 catalogs at the bright end,
errors in proper motion measurements due to field crowding also
possibly result in stars having overestimated proper motions, which
would in turn fall within the reduced proper motion range of M
dwarfs. Close examination of a number of stars at low Galactic
latitudes indeed raises doubt about the accuracy of the proper motion
quoted in HIPPARCOS and TYCHO-2 in at least a few cases.

The distribution of fainter ($V>12$) M dwarfs, on the other hand,
shows a net deficit of objects in the southern sky, especially for
$\delta\lesssim-30$. As discussed in \S2.3 this is due to the limited
coverage of SUPERBLINK in the south, where the survey is only complete
for proper motions $\mu>150$ mas yr$^{-1}$, compared to $\mu>40$ mas
yr$^{-1}$ in the north. The distribution, however, does not show any
overdensity/deficit of stars at low Galactic latitudes, which
suggests both a high completeness and low contamination from background
giants. This demonstrates the high efficiency and quality controls of
the SUPERBLINK survey in high density fields.  

\begin{figure*}[t]
\epsscale{1.10}
\plotone{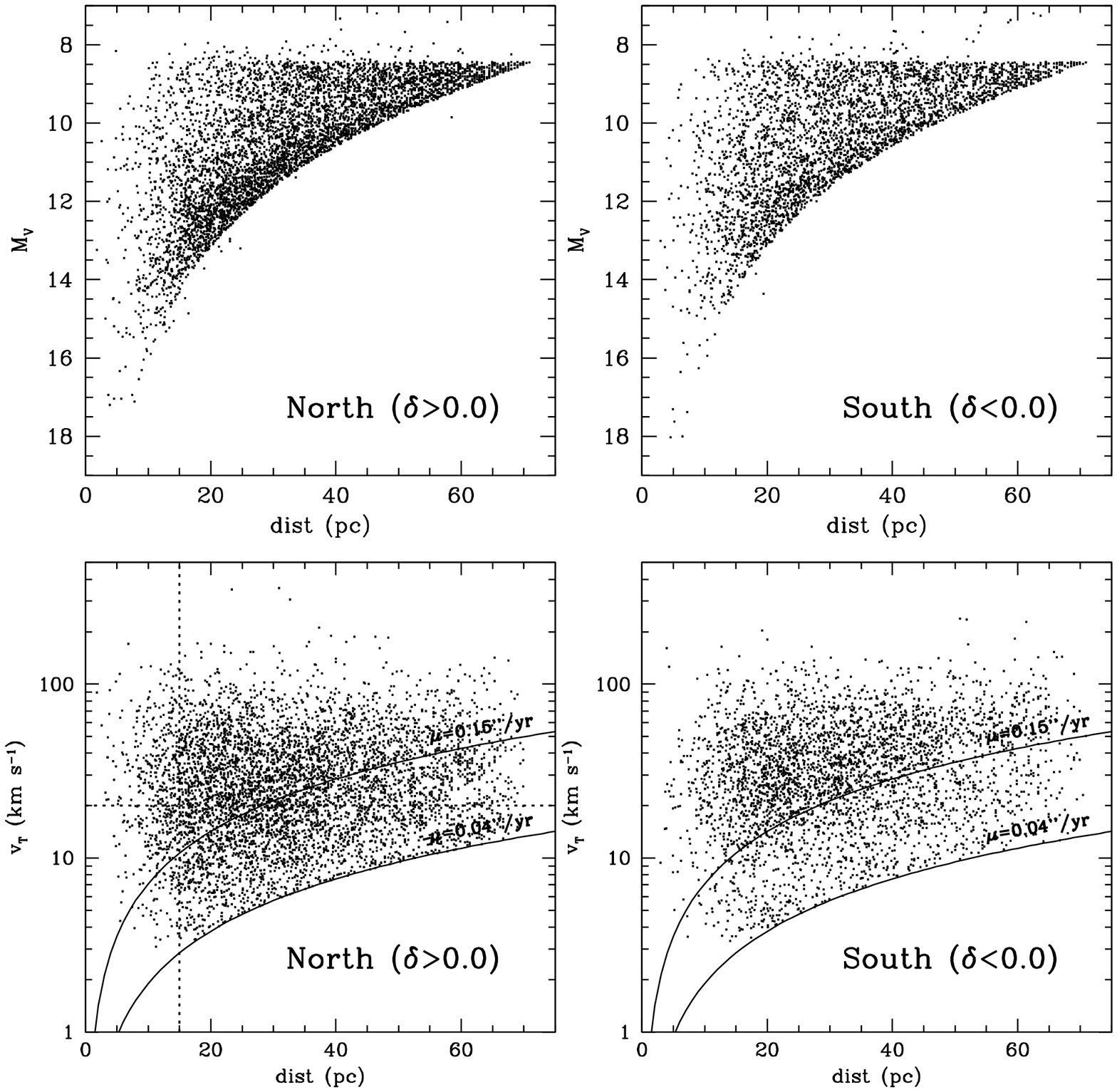}
\caption{Selection effects on the sample of bright M dwarfs. Top:
  absolute magnitude as a function of distance. Trigonometric
  parallaxes are used when available, photometric distances
  otherwise. The plot shows the effect of the limiting apparent
  magnitude ($J<10.0$). Bottom: transverse velocity as
  a function of distance, showing the kinematic bias due to the
  proper motion limit of the SUPERBLINK catalog ($\mu>40$ mas
  yr$^{-1}$ in the northern sky and parts of the south, and $\mu>150$
  mas yr$^{-1}$ elsewhere). The dashed lines in the lower-left panel
  show the range limits (v$_t>$20 km s$^{-1}$; d$<$15 pc) used to
  infer the distance and transverse velocity distributions, unbiased
  by kinematic selection effects (see Figure 12.)}
\end{figure*}

\begin{figure}[t]
\epsscale{2.20}
\plotone{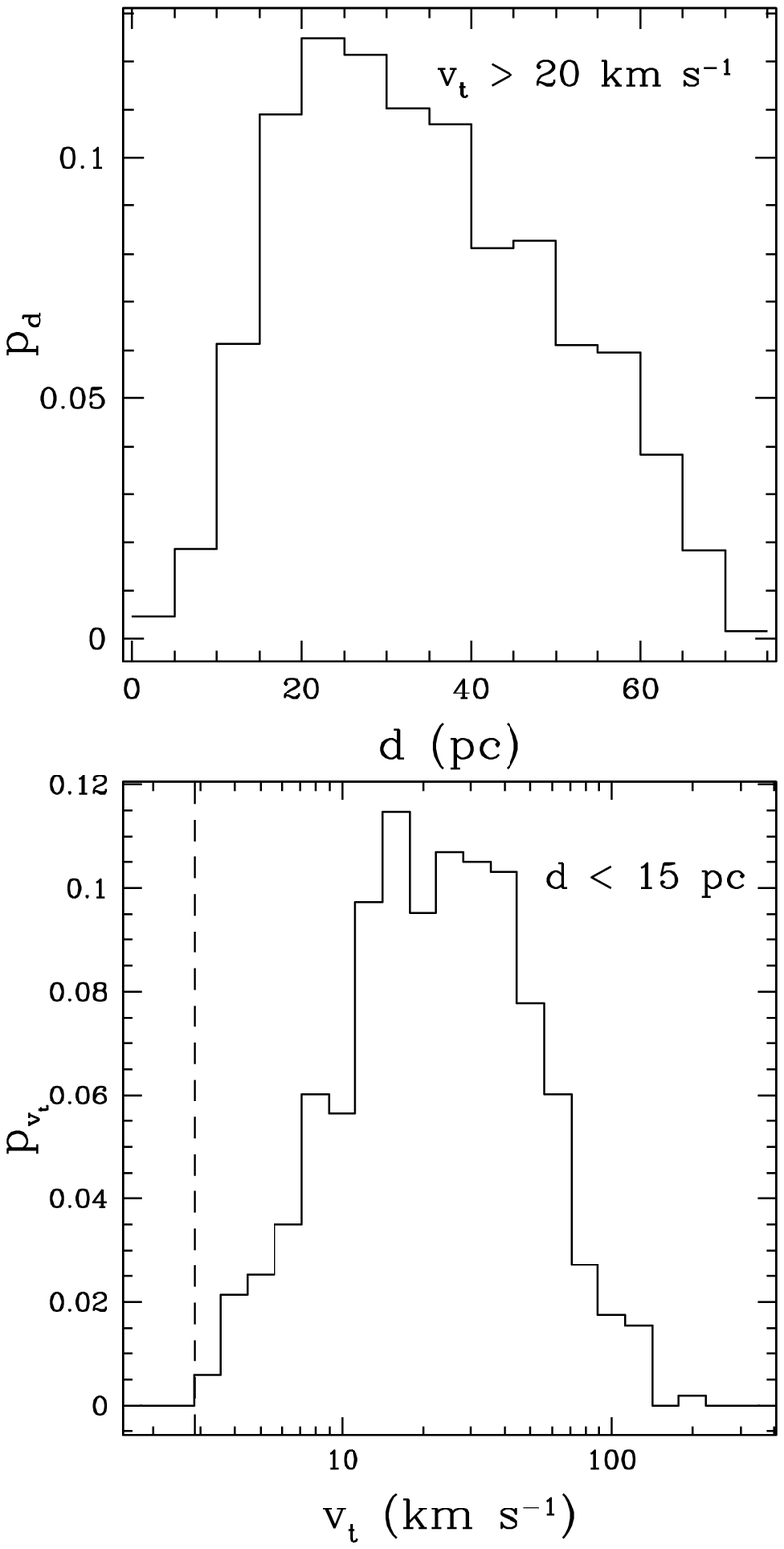}
\caption{Estimated statistical distributions of M dwarfs with proper
  motion $\mu>40$ mas yr$^{-1}$ and magnitude $J<10$, in the volume
  centered on the Sun. Top: distribution as a function of distance,
  evaluated from stars with transverse velocity $v_t>$20 km s$^{-1}$
  in the northern sky, where the census is believed complete
  (normalized). The sharp rise from 0 to 20pc is due to the geometric
  $p_d\sim d^{2}$ increase from uniform space density. The drop beyond
  d=20pc is due to both kinematic selection and the magnitude limit of
  the catalog. Bottom: distribution as a function of transverse
  velocity, estimated from  stars with distance d$<15$ pc from the
  northern sample. The dashed line shows the boundary above which the
  survey is expected to be complete, given the proper motion limits.}
\end{figure}

\subsection{Kinematic bias}

We illustrate in Figure 12 the effects of kinematic bias and magnitude
limit in our catalog. The upper panels show the estimated absolute
magnitudes as a function of distance, while the bottom panels show the
transverse motions $v_t$ (i.e. projected motions in the plane of the
sky) as a function of distance, for all the stars in our catalog. The
astrometric distances are used when available, otherwise we use the
photometric distances as calculated in \S3.4.

One cannot speak of incompleteness due to the magnitude limit, since
the $J<10$ magnitude limit was {\em imposed} as a primary constraint
to the catalog. Figure 12 however illustrates the effects of this
magnitude limit in restricting the distance range of the stars in the
catalog. All the M dwarfs with $J<10$ populate the nearest 75
parsecs. The limiting distance is also a function of absolute
magnitude thus spectral subtype. Late-type M dwarfs, which have lower
absolute magnitudes $M_V$, have a much shorter distance limit in our
catalog, which explains the strong deficit of stars with subtypes
later than M4 (see Figure 10).

The main source of incompleteness in our magnitude-limited catalog is
kinematic bias. Because all of our stars are selected among
sources with proper motions $\mu>40$ mas yr$^{-1}$, nearby stars with
low components of their transverse motion $v_t$ (i.e. projected motion
in the plane of the sky) may be missed. Nearby stars can have low
values of $v_t$ either because of a low intrinsic motion relative to
the Sun, or because their motion vector is nearly perpendicular to the
plane of the sky. For example, a star at 50 parsecs from the Sun with
$\mu<40$ mas yr$^{-1}$ has a transverse velocity $v_t<$9.48 km
s$^{-1}$, which will occur with a probability of about 10\% for stars
in the Solar neighborhood (see \S2.2 and Fig.1).

The SUPERBLINK catalog currently has a higher proper motion limit
($\mu<150$ mas yr$^{-1}$) in large swaths of the southern sky, which
makes our census much more prone to a kinematic bias. This explains
the deficit of sources in the Southern sky (Fig.11). This is only
mitigated at the bright end because of our use of the Hipparcos and
TYCHO-2 catalogs.

To estimate the overall completeness of our catalog, we proceed in two
steps. First we use Monte-Caro simulations to estimate the
completeness of the northern sky census, which we assume to suffer
only from kinematic bias due to the $\mu<40$ mas yr$^{-1}$ proper
motion limit. Then we assume that the local distribution of field
stars is isotropic in the vicinity of the Sun, and thus infer how many
stars are missing from the southern sky census, i.e. assuming that
both the north and south hemisphere should have the same number of
bright M dwarfs.

We first assume that the transverse velocity $v_t$ is independent of
the distance $d$ for stars in the Solar Neighborhood, and assume a
statistical distribution of the form
$p(v_t,d)=p_{v_t}(v_t)p_d=p_d(d)$. We extract from our data the
statistical distributions $p_{v_t}=p_{v_t}(v_t)$ and $p_d=p_d(d)$ from
appropriate cuts in $d$ and $v_t$. For $p_d$, we consider only stars in
the northern sky with transverse velocities $v_t>20$ km s$^{-1}$; this
boundary is illustrated in Figure 12 (lower-left panel), which shows
that the stars that satisfy this constraint do not suffer from the
kinematic bias, because all the stars in our catalog which have
$v_t>20$ km s$^{-1}$ also have proper motions $\mu>40$ mas
yr$^{-1}$. The normalized distribution function $p_d$ is shown in
Figure 13 (top panel). For $p_{v_t}$, we consider only stars with
distances $d<15$ parsecs; again the boundary is illustrated in Figure 12,
and shows that these very nearby stars suffer from minimal kinematic
bias, because stars with proper motions $\mu<40$ mas yr$^{-1}$ (and
missing from the census) can only have $v_t<2.8$ km s$^{-1}$, which
excludes very few stars. The normalized distribution function
$p_{v_t}$ is shown in Figure 13 (bottom panel) where the limit of
completeness ($v=2.8$ km s$^{-1}$) is shown for reference. Given the
sharp drop in $p_{v_t}$ for transverse motions $v_t<10$ km s$^{-1}$,
it is safe to assume that the sample of $d<15$ parsecs stars is
essentially complete.

We perform Monte-Carlo simulations based on the statistical
distribution $p_d$ and $p_{v_t}$ to estimate the probability that an M
dwarf with $J<10$ will be detected in our survey. We thus estimate
that our census of the northern sky has a completeness of
$91.8\pm0.3$\%. This means that the 5,361 stars in the northern sky
sample are most likely drawn from a hypothetical sample of
5,840$\pm$18 stars. The $\approx480$ stars missing due to the
kinematic bias are expected to have proper motions $\mu<40$ mas
yr$^{-1}$.

The above analysis assumes that the SUPERBLINK survey is complete
within the proper motion range and magnitude limit. If the survey
itself is incomplete, then the number of missing objects would be
larger. The completeness of the SUPERBLINK survey generally depends
magnitude and field density. However in the magnitude range under
consideration (8$<V<$15) a completeness test based on the fraction
of TYCHO-2 stars recovered by SUPERBLINK {\em and} the fraction of
SUPERBLINK stars listed in TYCHO-2, suggests that the combined
TYCHO-2/SUPERBLINK catalog has a completeness $>98\%$
\citep{Lepine.Shara.2005}. If we factor in the possibility that
$\approx$2\% stars in that magnitude range have escaped detection by
both TYCHO-2 and SUPERBLINK, this would bring the total number of
hypothetical bright M dwarfs in the northern sky to $\approx$5,950
stars. 

\subsection{Southern sky incompleteness}

Our southern sky census is significantly less complete than the
northern census. While our catalog has 5,361 stars north of the celestial
equator, it only has 3,528 in the south. The main reason for the
incompleteness is the limited coverage of the SUPERBLINK proper motion
survey in the south. The survey is complete to $\mu>40$ mas yr$^{-1}$
for the entire northern sky, but reaches that limit only over a
fraction of the south, mainly in the declination range
$-20<\delta<0$ (see Fig.11) South of $\delta=-20$, the survey is only
complete to $\mu>150$ mas yr$^{-1}$, but reaches to $\mu>80$ mas
yr$^{-1}$ in some areas. The incompleteness of the SUPERBLINK survey
is mitigated by the inclusion of the TYCHO-2 catalog, which is mostly limited by
magnitude and actually includes stars with proper motions smaller than
the SUPERBLINK limit of $\mu<$40 mas yr$^{-1}$. However the magnitude
limit of the TYCHO-2 catalog, $V\lesssim$12.0, is too bright to
include all but a fraction of the M dwarfs in the magnitude range of
interest ($J<10.0$).

Under the assumption that the northern sky census is fully
representative of the entire volume around the Sun, i.e. assuming the
M dwarf distribution to be {\em isotropic} in the Solar neighborhood,
we use the results from the Monte-Carlo simulation above and estimate
that the southern-sky census of M dwarfs brighter than $J=10$ should
also include 5,950 stars, for a total of 11,900 M dwarfs over the
entire sky. Given that our southern sample includes only 3,528 stars,
we thus conclude that we are currently missing $\approx$2,420 stars in
the southern sky. In other words our southern sky census has a
completeness of only $\approx$59.3\%.

\section{Discussion and conclusions}

The need to assemble a comprehensive list of {\em bright} M dwarfs
over the entire sky is primarily motivated by the search for
extra-solar planets. Based on current techniques, M dwarfs are the most
promising stars around which to look for Earth-mass planet in a
habitable zone. Radial velocity measurements are most sensitive to the
reflex motion from planets in close orbits around their parent stars,
or to planets orbiting low-mass stars; for planets in habitable zones,
this favors M dwarfs, whose zone is closer in, and whose masses
are relatively low. Transit surveys have a higher chance of
detecting planets in close orbits, and are more sensitive to planets
orbiting smaller stars; this likewise favors M dwarfs as the most
promising targets. The main caveat is that both radial velocity
measurements and transit monitoring are most efficient for relatively
bright stars. With current observational resources, radial-velocity
monitoring can realistically be performed on stars with absolute
magnitudes $V<13.0$, and near-term developments only promise to
increase the depth of surveys by one or two magnitudes, most notably
by observing at infrared wavelengths where M dwarfs are noticeably
brighter. M dwarfs with apparent infrared magnitudes $J<10$ will soon
be within range of radial velocity surveys.

The motivation of the SUPERBLINK proper motion survey is to identify
the largest possible number of hydrogen-burning objects up to at least
100 parsecs of the Sun. With a low proper motion limit of $\mu>40$ mas
yr$^{-1}$ and astrometric accuracy $8$ mas yr$^{-1}$, the catalog
suffers from minimal kinematic bias to a distance of 50-60 parsecs,
which is the range over which one would expect to find M dwarfs with
$J<10$. The current SUPERBLINK catalog is complete over the entire
northern sky, but the southern sky survey is still in progress. At
this time, the catalog has a proper motion limit $\mu>40$ mas
yr$^{-1}$ in the declination range $-20<\delta<+90$; at southern
declinations the proper motion limit is now $\mu>150$ mas yr$^{-1}$,
except for relatively bright stars ($V<12$) for which the catalog
extends to $\mu>40$ mas yr$^{-1}$ because of the inclusion of data
from the TYCHO-2 and HIPPARCOS catalogs.

Our search of the current version of the SUPERBLINK catalog
has identified 8,889 bright M dwarfs, of which 5,361 are in the northern
sky and 3,528 in the south. The lower number of stars catalogued in
the south reflects the limited extent of our proper motion survey in
the south. Our estimate of the kinematic bias suggests that our
northern census is $91.8\pm0.3$\% complete. From this we infer that
there should be $\approx11,900$ M dwarfs with magnitudes $J<10$ over
the entire sky. Our catalog currently lists $\approx76$\% of the
entire set. We recommend that these stars be observed by
radial-velocity monitoring programs with the highest priority.
In addition to exoplanet surveys, completing the census of the Solar
neighborhood is important to constrain the stellar density in the
Galaxy, and to determine the distribution of stellar properties. Our
catalog includes many stars which are identified here for the first
time as members of the Solar neighborhood. In particular, most of the
sources are not listed in the Catalog of Nearby Stars.

The key to completing the census of bright M dwarfs in the south is to
obtain a proper motion catalog with a lower proper motion limit,
currently limited to $\mu>150$ mas yr$^{-1}$ south of $\delta=-20$.
We are currently working on an extension of the SUPERBLINK proper
motion survey, which will extend the limits to $\mu>40$ mas yr$^{-1}$
to $\delta<-30$, and to $\mu>80$ mas yr$^{-1}$ in the $-30<\delta<-90$
range. We predict that extended surveys will identify an additional
$\approx2,700$ bright M dwarfs, if a proper motion limit $\mu>40$ mas
yr$^{-1}$ can be reached over the entire sky. A revised catalog will
be produced as the survey reaches completion.

At this time we are conducting a systematic spectroscopic follow-up
survey of the brightest stars in this catalog, and searching the
literature for previous determinations of spectral subtypes. Results
from these investigations will be provided in upcoming papers by our
team.

\acknowledgements

This material is based upon work supported by the National Science
Foundation under Grants No. AST 06-07757, AST 09-08419, and AST
09-08406.


\end{document}